# Tangential contacts of three-dimensional power-law graded elastic solids: A general theory and application to partial slip

## Markus Heß[a],*, Qiang Li[a]

[a]*Department of System Dynamics and Friction Physics, Technische Universität Berlin, Sekr. C8-4, Straße des 17. Juni 135, 10623 Berlin*

**Abstract**

A rigorous theory for solving tangential contacts between three-dimensional power-law graded elastic solids of arbitrary geometry is presented. For multiple contacts such as those occurring between two nominally flat but rough half-spaces, the well-known Ciavarella-Jäger theorem is established accompanied by a discussion of tangential coupling. Nevertheless, the focus of the work is on axisymmetric single contacts under arbitrary unidirectional tangential loading, for which closed-form analytical solutions are derived based on the Mossakovskii-Jäger procedure. In comparison to the results of common approximate methods, the solutions include the non-axisymmetric components of tangential displacements, which are indispensable for the accurate determination of the relative slip components and thus the surface density of frictional energy dissipation in the partial slip regime. Although a simplified approach is used for the calculation of the dissipated energy density, the results in the limiting case of homogeneous material are in excellent agreement with those from a full numerical computation. As an application example, the complete solutions for the tangential contact of parabolically shaped power-law graded elastic solids in the partial slip regime are derived and the influence of the material gradient as well as Poisson's ratio on the surface density of dissipated energy is investigated.

*Keywords:* Contact mechanics; Functionally graded materials; Fretting; Partial slip; Frictional energy dissipation; Crack nucleation; Ciavarella-Jäger theorem

## 1. Introduction

Compared to conventional composites, functionally graded materials (FGMs) are characterized by a gradually changing composition or microstructure according to a predefined law. In this way, controlled gradients in the thermal, electrical, or mechanical properties can be induced to meet specific component requirements, such as high hardness on the outside coupled with high ductility in the interior. The field of practical application of FGMs is vast and ever-growing [1]. For instance, in mechanical engineering FGMs are used for numerous components such as cutting tools, gears, engine pistons, combustion chambers, turbine blades or bearing liners since they improve the strength, thermal or corrosion resistance. In addition, biomedical applications should be highlighted including artificial joints, whose biocompatibility and wear resistance are significantly improved by using FGMs, thus extending the service life of endoprostheses [2]. The continuously varying volume fractions of the material constituents of FGMs are usually associated with a continuously changing Young's modulus in the direction of grading, which makes the solution of contact problems extremely difficult and typically requires numerical methods [3]. In a recent publication, Argatov and Sabina [4] have presented a method to recover information on the in-depth grading of elastic FGMs by means of indentation experiments. The applicability of their method is demonstrated by numerous examples of shear modulus variation with depth. However, although already in the 50 years before the origin of FGMs the influence of elastic inhomogeneities on stresses and displacements within the half-space was intensively studied in the field of soil mechanics (see e.g. [5], [6]), closed-form analytical solutions of contact problems are known only for a few special laws of in-depth grading. These include the power law and to some extent the exponential one as well. The great advantage of the power law according to

$$E(z) = E_0 \left( \frac{z}{c_0} \right)^k \quad \text{with} \quad -1 < k < 1 \ , \tag{1}$$

where $c_0$ denotes a characteristic depth, is that the surface displacements caused by a (arbitrarily directed) point force on the half-space are known [7]. Although almost all works on power-law graded materials presuppose a positive exponent $k$, the Green's functions derived by Booker et al. [7] apply to negative exponents as well [8] [9]. Fig. 1 depicts the qualitative gradients for negative and positive exponents of elastic inhomogeneity as part of a tangential contact with a rigid, convex indenter.

* Corresponding author. Tel.: +4903031421485.
  *E-mail address:* markus.hess@tu-berlin.de

For negative exponents, Young's modulus decreases with depth, starting from an (infinitely) stiff surface to an (infinitely) soft core (Fig. 1a). For positive exponents the inverse characteristic holds (Fig. 1c). Although the limit values of zero and infinity contradict real material behavior, the solutions can reflect the tendencies of real material behavior and thus contribute significantly to understanding.

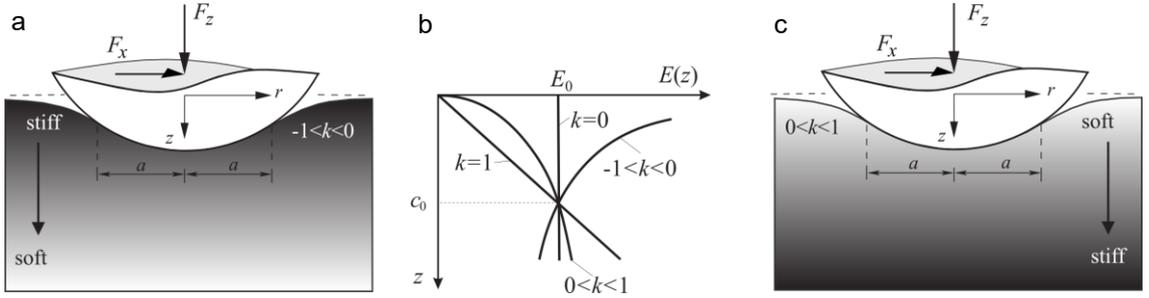

Fig. 1. Tangential contact of a rigid convex indenter and a power-law-graded half-space whose modulus (a) decreases with depth and (c) increases with depth; (b) illustrates the dependence of Young's modulus on the depth coordinate for different exponents of elastic inhomogeneity.

Frictionless normal contact problems between rigid indenters and a power-law graded, elastic half-space have been solved primarily by Booker et al. in 1985 [10] and Giannakopoulos and Suresh in 1997 [11]. Surprisingly, it took more than 10 years until the solutions were extended to adhesive normal contacts which is probably because of the integrals to be solved, which are still very complex. Solutions of adhesive contacts of power-law graded solids have been mainly presented by Jin, Guo, Gao and co-workers in a series of publications [12] [13] [14]. Quite recently, they developed a generalized comprehensive Maugis-Dugdale solution [15]. A simple method for solving normal contacts between axisymmetric power-law graded elastic solids considering adhesion according to Johnson, Kendall and Roberts was developed by Heß [16]. Argatov et al. [17] extended the method for application to arbitrary laws of in-depth grading. At a 2016 conference [18], Heß presented a method for solving tangential contacts of power-law graded, elastic solids including full-stick, partial slip, and gross sliding. However, immediately after the conference, only some final results were published without proof to give collaborators and further researchers quick access to the mapping rules in the context of the method of dimensionality reduction (MDR) [19] (see also [20]). Therefore, one of the authors' motivations for the present work is to provide the complete derivation of the solutions. Another motivation involves the development of a more general theory for solving axisymmetric normal and tangential contacts under arbitrary boundary conditions, which additionally allows to precisely determine the non-axisymmetric parts of tangential surface displacements. This is important for tasks that require detailed information about micro-slip. For instance, we will show that the surface density of energy dissipation in the partial slip regime can be approximated very well when considering the non-axisymmetric component of micro-slip. Various contact mechanics methods make use of the so-called Mindlin approximation, i.e., the magnitude of relative slip is assumed to be axisymmetrically distributed. If one is interested in the total energy dissipated in a full cycle of tangential loading, this assumption is justified, but it is not suitable for capturing the surface density of energy dissipation [21]. However, the latter can be relevant for predicting the location of surface damage and crack nucleation [22].

In addition to the general theory for solving axisymmetric normal and tangential contacts of power-law graded elastic solids, the integral equations for handling multiple contacts like those between two nominally flat but rough surfaces are also discussed with coupling effects taken into account. In this context the well-known Ciavarella-Jäger theorem [23] is adapted to be applied to power-law graded, elastic solids.

The paper is organized as follows. Chapter 2 summarizes the mathematical fundamentals of locally isotropic power-law graded elastic half-spaces. Based on Mossakovskii's principle of superposition [24], a general solution for axisymmetric normal loading is derived in Chapter 3. All relevant quantities depend only on a single function, which must be determined from the boundary conditions. It is distinguished between the prescription of an axisymmetric pressure distribution on a circular area and an indentation problem. In an analogous way (inspired by a work of Jäger [25]), a general solution for uni-directional tangential tractions (axisymmetric in magnitude) applied to a circular area is developed in Chapter 4. It requires the solutions for a rigid-body translation of the entire loading area in tangential direction. The corresponding very complex calculations have been performed in the Appendix. In Sect. 4.3, the theory is applied to solve the problem of a sphere sliding on a power-law graded half-space. Chapter 5 first deals with the case where two power-law graded solids (but with the same exponent of elastic inhomogeneity) make contact over an arbitrarily shaped area. The conditions for complete decoupling of normal and tangential effects as well as between the components of the tangential stresses themselves are addressed in detail. Under complete decoupling the famous Ciavarella-Jäger theorem for multiple contacts is formulated und a sound approximation for the case of tangential coupling according to Putignano et al. [26] is added. The same applies to partial slip between axisymmetric bodies in the presence of tangential coupling, for which simple calculation equations for the relative slip components are established as well. Finally, the presented theory is used to solve the normal and partial slip contact of a sphere and a half-space made of elastically similar power-law graded material. Both the relative slip components and the energy density are determined and the influence of the exponent of elastic inhomogeneity as well as Poisson's ratio is investigated.

## 2. Fundamentals of a power-law graded elastic half-space

### 2.1. Green's functions for point loading

For a point force $\boldsymbol{F} := \left\{ F_x, F_y, F_z \right\}$ acting at the origin on the surface of a power-law graded elastic half-space $z \geq 0$, the surface displacements of a point (x,y,0) were determined by Booker et al. [7]

$$\bar{u}_x = \frac{c_0^k}{E_0 r^{3+k}} \left[ \left( Hx^2 + Ky^2 \right) F_x + \left( H - K \right) xy F_y - Lxr F_z \right]$$

$$\bar{u}_y = \frac{c_0^k}{E_0 r^{3+k}} \left[ \left( H - K \right) xy F_x + \left( Kx^2 + Hy^2 \right) F_y - Lyr F_z \right],$$

$$\bar{u}_z = \frac{c_0^k}{E_0 r^{3+k}} \left[ Lxr F_x + Lyr F_y + Br^2 F_z \right]$$

(2)

where $r = \sqrt{x^2 + y^2}$ and the coefficients $H$, $K$, $L$ and $B$ depend on both the exponent of the elastic inhomogeneity and the Poisson's ratio according to

$$B = \left(1 - \nu^2\right) \frac{\sin\left(\frac{\beta\pi}{2}\right)\beta}{\left(1+k\right)^2}\frac{\Gamma\left(\frac{3+k+\beta}{2}\right)\Gamma\left(\frac{3+k-\beta}{2}\right)}{\Gamma\left(1+\frac{k}{2}\right)^2}, \quad H = \frac{1}{\pi\left(1-k\right)}\left[1 + \nu - \left(1 - \nu^2\right)\frac{\sin\left(\frac{\beta\pi}{2}\right)k}{\beta}\frac{\Gamma\left(\frac{3+k+\beta}{2}\right)\Gamma\left(\frac{3+k-\beta}{2}\right)}{\Gamma\left(1+\frac{k}{2}\right)^2}\right]$$

$$L = -\left(1 - \nu^2\right)\frac{\cos\left(\frac{\beta\pi}{2}\right)}{k\pi}\frac{\Gamma\left(\frac{3+k+\beta}{2}\right)\Gamma\left(\frac{3+k-\beta}{2}\right)}{\Gamma\left(\frac{1+k}{2}\right)\Gamma\left(\frac{3+k}{2}\right)}, \quad K = \frac{1}{\pi\left(1-k\right)}\left[\left(1 - \nu^2\right)\frac{\sin\left(\frac{\beta\pi}{2}\right)}{\beta}\frac{\Gamma\left(\frac{3+k+\beta}{2}\right)\Gamma\left(\frac{3+k-\beta}{2}\right)}{\Gamma\left(1+\frac{k}{2}\right)^2} - \left(1 + \nu\right)k\right] \tag{3}$$

with $\beta = \sqrt{\left(1 - \frac{k\nu}{1-\nu}\right)\left(1+k\right)}$. The limiting case $k = 0$ represents elastically homogeneous material and the coefficients in Eq. (3) simplify to:

$$B = K = \frac{1-\nu^2}{\pi}, \quad H = \frac{1+\nu}{\pi}, \quad L = \frac{\left(1+\nu\right)\left(1-2\nu\right)}{2\pi}. \tag{4}$$

*2.2. Loading by distributed normal and tangential stresses*

The surface displacements caused by normal pressure $p$ and tangential tractions $q_x$ and $q_y$ distributed over an area $A$ of the surface are obtained by superposition of solutions of the type given by Eq. (2)

$$\bar{u}_x\left(x,y\right) = \frac{c_0^k}{E_0}\int_A\left(\frac{K}{s^{1+k}}q_x + \left(H - K\right)\left(\frac{\left(x-\tilde{x}\right)^2}{s^{3+k}}q_x + \frac{\left(x-\tilde{x}\right)\left(y-\tilde{y}\right)}{s^{3+k}}q_y\right) - L\frac{x-\tilde{x}}{s^{2+k}}p\right)dA$$

$$\bar{u}_y\left(x,y\right) = \frac{c_0^k}{E_0}\int_A\left(\left(H - K\right)\left(\frac{\left(x-\tilde{x}\right)\left(y-\tilde{y}\right)}{s^{3+k}}q_x + \frac{\left(y-\tilde{y}\right)^2}{s^{3+k}}q_y\right) + \frac{K}{s^{1+k}}q_y - L\frac{y-\tilde{y}}{s^{2+k}}p\right)dA, \tag{5}$$

$$\bar{u}_z\left(x,y\right) = \frac{c_0^k}{E_0}\int_A\left(L\left(\frac{x-\tilde{x}}{s^{2+k}}q_x + \frac{y-\tilde{y}}{s^{2+k}}q_y\right) + \frac{B}{s^{1+k}}p\right)dA$$

where $s = \sqrt{\left(x-\tilde{x}\right)^2 + \left(y-\tilde{y}\right)^2}$.

# 3. Normal loading over a circular area

*3.1. General solution for axisymmetric normal loading*

It is clear and immediately evident from Eqs. (5) that even a pure normal loading generally causes tangential displacements at the surface. However, with a foresight to the later to be investigated tangential contact between two power-law graded elastic bodies under complete decoupling condition, our interest is only directed to the normal displacements of the surface. For a pressure distribution

$$p\left(r\right) = \frac{p_0 a^{1-k}}{\left(a^2 - r^2\right)^{\frac{1-k}{2}}} \tag{6}$$

applied to a circular region of radius $a$, the surface normal displacements read [10]

$$\bar{u}_z\left(r\right) = \begin{cases} \dfrac{p_0\pi^2 a}{\cos\left(\frac{k\pi}{2}\right)}\dfrac{B}{E_0}\left(\dfrac{c_0}{a}\right)^k & \text{for } r \leq a \\[4mm] \dfrac{2\pi p_0 a}{1+k}\dfrac{B}{E_0}\left(\dfrac{c_0}{a}\right)^k\left(\dfrac{a}{r}\right)^{1+k}{}_2\text{F}_1\left(\dfrac{1+k}{2},\dfrac{1+k}{2};\dfrac{3+k}{2};\dfrac{a^2}{r^2}\right) & \text{for } r \geq a \end{cases}, \tag{7}$$

where $_2F_1(...)$ denotes the Gaussian hypergeometric function. Since the displacements of all points in the loading area are equal, the pressure given by Eq. (6) is exactly that which would result from indentation by a flat-ended cylindrical rigid punch of radius $a$. The normal force is obtained by integration of the pressure distribution (6) over the loaded region, which leads to

$$F_z = \frac{2\pi p_0 a^2}{k+1}. \tag{8}$$

To solve arbitrary axisymmetric problems, a procedure developed by Mossakovskii [24] and especially revitalized by Jäger [25] can be applied. Imagine that an arbitrary axisymmetric normal displacement on a circle of radius $a$ at the surface is prescribed. Then it is easy to recognize that this displacement can also be generated by superposition of (infinitesimal) indentations of flat-ended rigid punches having different radii. This automatically means that all quantities can be obtained by superposition of flat-ended punch solutions. What has been said applies equally if instead of the displacements an arbitrary axisymmetric pressure distribution on the circular area is given. Using Eqs. (6) - (8), the differential contributions of a flat-ended rigid punch of radius $\tilde{a}$ to normal stress, displacement and normal force can be specified as follows:

$$dp(r;\tilde{a}) = \frac{\tilde{a}^{1-k}}{\left(\tilde{a}^2 - r^2\right)^{\frac{1-k}{2}}} p_0{}'(\tilde{a})d\tilde{a} \quad \text{for } r \leq \tilde{a}, \tag{9}$$

$$d\overline{u}_z(r;\tilde{a}) = \begin{cases} \dfrac{\pi^2 \tilde{a}}{\cos\left(\frac{k\pi}{2}\right)} \dfrac{B}{E_0} \left(\dfrac{c_0}{\tilde{a}}\right)^k p_0{}'(\tilde{a})d\tilde{a} & \text{for } r \leq \tilde{a} \\[3ex] \dfrac{2\pi\tilde{a}}{1+k} \dfrac{B}{E_0} \left(\dfrac{c_0}{\tilde{a}}\right)^k \left(\dfrac{\tilde{a}}{r}\right)^{1+k} {}_2F_1\left(\dfrac{1+k}{2}, \dfrac{1+k}{2}; \dfrac{3+k}{2}; \dfrac{\tilde{a}^2}{r^2}\right) p_0{}'(\tilde{a})d\tilde{a} & \text{for } r \geq \tilde{a} \end{cases}, \tag{10}$$

$$dF_z(\tilde{a}) = \frac{2\pi\tilde{a}^2}{k+1} p_0{}'(\tilde{a})d\tilde{a}. \tag{11}$$

Only flat punches with radii $r \leq \tilde{a} \leq a$ contribute to the normal stress at location $r \leq a$. Hence it follows

$$p(r) = \int_r^a \frac{\tilde{a}^{1-k}}{\left(\tilde{a}^2 - r^2\right)^{\frac{1-k}{2}}} p_0{}'(\tilde{a})d\tilde{a} = \int_r^a \frac{n_z(\tilde{a})}{\left(\tilde{a}^2 - r^2\right)^{\frac{1-k}{2}}} d\tilde{a} \quad \text{with } n_z(\tilde{a}) = \tilde{a}^{1-k} p_0{}'(\tilde{a}) \tag{12}$$

where we have introduced the function $n_z(\tilde{a})$. When setting up the formula for the displacement, it should be noted that points located at $0 < r < a$ are outside the respective loading areas of flat punches with radii $\tilde{a} < r$, and inside those with radii $r < \tilde{a} < a$. Hence, both parts mentioned in Eq. (10) contribute, while the displacement of points located at $r > a$ can be calculated by a pure superposition of contributions of the kind $\tilde{a} < r$. Both cases can be captured by a single equation:

$$\overline{u}_z(r) = \frac{2\pi}{1+k} \frac{Bc_0{}^k}{E_0} \int_0^{\text{Min}\{r,a\}} \left(\frac{\tilde{a}}{r}\right)^{1+k} {}_2F_1\left(\frac{1+k}{2}, \frac{1+k}{2}; \frac{3+k}{2}; \frac{\tilde{a}^2}{r^2}\right) n_z(\tilde{a})d\tilde{a} + \frac{\pi^2}{\cos\left(\frac{k\pi}{2}\right)} \frac{Bc_0{}^k}{E_0} \int_{\text{Min}\{r,a\}}^a n_z(\tilde{a})d\tilde{a}. \tag{13}$$

For a point outside of the loading area, $\text{Min}\{r,a\} = a$ holds and the second integral becomes zero. The total normal force is obtained by superposition of incremental forces (11)

$$F_z = \frac{2\pi}{k+1} \int_0^a \tilde{a}^{1+k} n_z(\tilde{a})d\tilde{a}. \tag{14}$$

By means of Eqs. (12) to (14) the solutions for arbitrary axisymmetric normal loading on a circular domain are found. They only depend on the single function $n_z(\tilde{a})$, which has to be determined from the boundary conditions.

### 3.1.1. Case I: Prescribed pressure applied to a circular region

Let us first assume that an axisymmetric pressure distribution is prescribed over a circular area of radius $r$ on a power-law graded elastic half-space

$$\sigma_{zz}(r) = \begin{cases} -p(r) & \text{for } r \leq a \\ 0 & \text{for } r > a \end{cases} \tag{15}$$

and no shear stresses act on it. In this case, the sought function $n_z(\tilde{a})$ must be determined from Eq. (12), according to which the pressure is expressed as an Abel transformation of $n_z(\tilde{a})$. The corresponding inverse Abel transformation leads to

$$n_z(\tilde{a}) = -\frac{2\cos\left(\frac{k\pi}{2}\right)}{\pi} \frac{d}{d\tilde{a}} \int_{\tilde{a}}^{a} \frac{r p(r)}{\left(r^2 - \tilde{a}^2\right)^{\frac{1+k}{2}}} dr \quad \text{for } 0 < \tilde{a} < a . \tag{16}$$

Substituting Eq. (16) into (13) and (14) yields the normal displacements inside and outside the loaded area as well as the normal force. The latter can of course also be calculated by integrating the pressure over the loaded area. In any case, the elastic problem is completely solved.

### 3.1.2. Case II: Frictionless axisymmetric indentation (prescribed normal displacement)

In the case of indentation problems, mixed boundary conditions are present. Consider an axisymmetric rigid indenter whose shape is given by a function $f(r)$ and which is pressed into the elastically graded half-space by means of an externally applied normal force. The corresponding contact radius of the simply connected contact area is termed $a$ and the indentation depth $\delta_z$. Assuming frictionless contact, the mixed boundary conditions are as follows

$$\begin{aligned} \bar{u}_z(r) &= \delta_z - f(r) && \text{for } r \leq a \\ \sigma_{zz}(r) &= 0 && \text{for } r > a \end{aligned} . \tag{17}$$

Partial integration of Eq. (13) for $r \leq a$ and comparison with the displacement part of the boundary condition (17) gives

$$\delta_z = \frac{\pi^2}{\cos\left(\frac{k\pi}{2}\right)} \frac{Bc_0^k}{E_0} \int_0^a n_z(\tilde{a}) d\tilde{a} , \tag{18}$$

$$f(r) = 2\pi \frac{Bc_0^k}{E_0} \int_0^r \frac{\tilde{a}^k}{\left(r^2 - \tilde{a}^2\right)^{\frac{1+k}{2}}} \int_0^{\tilde{a}} n_z(s) ds \, d\tilde{a} . \tag{19}$$

Once again, Eq. (19) describes a generalized Abel transformation. Application of its inverse transformation and subsequent differentiation to $\tilde{a}$ yield

$$n_z(\tilde{a}) = \frac{E_0 \cos\left(\frac{k\pi}{2}\right)}{\pi^2 Bc_0^k} \frac{d}{d\tilde{a}} \left[ \frac{1}{\tilde{a}^k} \frac{d}{d\tilde{a}} \int_0^{\tilde{a}} \frac{r f(r)}{\left(\tilde{a}^2 - r^2\right)^{\frac{1-k}{2}}} dr \right] . \tag{20}$$

If the shape function of the indenter is known, $n_z(\tilde{a})$ can be determined according to (20). The normal displacements inside and outside the contact area, the normal force and the indentation depth are then easily obtained from Eqs. (13)

, (14) and (18). It should be noted that using a different parameterized function, alternative solutions for axisymmetric contacts can be found in [13], [16].

## 4. Tangential loading over a circular area

### 4.1. Uniform tangential displacement

The applicability of the superposition principle of Mossakovskii and Jäger is by no means limited to elastic normal contact problems. It can also be used to solve tangential or torsional contacts [25]. However, the basic prerequisite is that the solution for the corresponding rigid body motion is known. Thus, to find a solution for unidirectional tangential stresses of arbitrary axisymmetric distributed magnitude applied to a circular region on a power-law graded elastic half-space, the solution for a uniform tangential shift of the circular area is required. This solution was first presented at a conference in 2016 [18], but to date only parts of them and no derivation have been published [27], [19]. A complete derivation of the solution listed below can be found in Appendix A. .

A unidirectional tangential traction of the type

$$q_x(r) = \frac{q_0 a^{1-k}}{\left(a^2 - r^2\right)^{\frac{1-k}{2}}} \tag{21}$$

applied to a circular region of radius $a$ on the half-space causes tangential surface displacements in the direction of loading

$$\bar{u}_x(r,\theta) = \begin{cases} \dfrac{\pi^2 q_0 a^{1-k} c_0^k}{2\cos\left(\frac{k\pi}{2}\right)E_0}\left(H+K\right) & \text{for } r \le a \\[4mm] \dfrac{\pi q_0 a^{1-k} c_0^k}{(k+1)E_0}\left(\dfrac{a}{r}\right)^{1+k}\left[\left(H+K\right){}_2F_1\left(\dfrac{1+k}{2},\dfrac{1+k}{2};\dfrac{3+k}{2};\dfrac{a^2}{r^2}\right)+\left(H-K\right)\left(1-\dfrac{a^2}{r^2}\right)^{\frac{1-k}{2}}\cos 2\theta\right] & \text{for } r \ge a \end{cases} \tag{22}$$

and tangential displacements perpendicular to the loading

$$\bar{u}_y(r,\theta) = \begin{cases} 0 & \text{for } r \le a \\[3mm] \dfrac{\pi q_0 a^{1-k} c_0^k}{(k+1)E_0}\left(H-K\right)\left(\dfrac{a}{r}\right)^{1+k}\left(1-\dfrac{a^2}{r^2}\right)^{\frac{1-k}{2}}\sin 2\theta & \text{for } r \ge a \end{cases} \tag{23}$$

where $\theta$ represents the polar angle, hence $\cos 2\theta = \left(x^2 - y^2\right)/r^2$ and $\sin 2\theta = 2xy/r^2$. According to Eqs. (22) and (23), all points in the loaded area undergo the same displacement in the $x$-direction and no displacement in the $y$-direction. Thus, the entire loaded circle performs a rigid-body displacement in the direction of traction, denoted by $\bar{u}_{x,0}$. The tangential force is determined by integrating the tangential stresses (21) over the loaded area

$$F_x = \frac{2\pi q_0 a^2}{k+1}. \tag{24}$$

By relating the tangential force to the rigid-body displacement the tangential stiffness is defined by

$$k_x := \frac{F_x}{\bar{u}_{x,0}} = \frac{4\cos\left(\frac{k\pi}{2}\right)}{(1+k)\pi}\left(\frac{a}{c_0}\right)^k \frac{E_0 a}{H+K} \tag{25}$$

[27], which in the special case of homogeneous material provides the well-known result of Mindlin [28]

$$k_x = \frac{8Ga}{2-\nu} . \tag{26}$$

*4.2. General solution for uni-directional tangential tractions (axisymmetric in magnitude) applied to a circular area*

   To derive the equations for arbitrary uni-directional tangential stress distributions axisymmetric in magnitude, one can proceed completely analogously to the derivation in Section 3.1 for normal loading. The general solution can be thought as a superposition of differential rigid body displacements of different radii $\tilde{a}$. In this way, considering the results of the last section, the tangential stresses can be expressed as follows,

$$q_x(r) = \int_r^a \frac{n_x(\tilde{a})}{\left(\tilde{a}^2 - r^2\right)^{\frac{1-k}{2}}} d\tilde{a} , \tag{27}$$

where the unknown function $n_x(\tilde{a})$ will be determined later from the boundary conditions. Furthermore, the application of this special superposition principle leads to a tangential force, to be calculated from

$$F_x = \frac{2\pi}{k+1} \int_0^a \tilde{a}^{1+k} n_x(\tilde{a}) d\tilde{a} . \tag{28}$$

When calculating the tangential displacements, it should be noted once again that the displacements of points within the loading area $r < a$ are composed of different proportions, since they are located outside of the differential rigid shift areas with smaller radii and inside those with larger radii. Both, tangential displacements inside and outside of the loading area can be noted by one equation. The tangential displacements in the direction of loading are given by

$$\bar{u}_x(r,\theta) = \frac{\pi(H+K)c_0^{\,k}}{(1+k)E_0} \int_0^{\mathrm{Min}\{r,a\}} \left(\frac{\tilde{a}}{r}\right)^{1+k} {}_2F_1\left(\frac{1+k}{2}, \frac{1+k}{2}; \frac{3+k}{2}; \frac{\tilde{a}^2}{r^2}\right) n_x(\tilde{a}) d\tilde{a} + \frac{\pi^2(H+K)c_0^{\,k}}{2\cos\left(\frac{k\pi}{2}\right)E_0} \int_{\mathrm{Min}\{r,a\}}^a n_x(\tilde{a}) d\tilde{a}$$
$$+ \frac{\pi(H-K)c_0^{\,k}}{(1+k)E_0} \int_0^{\mathrm{Min}\{r,a\}} \left(\frac{\tilde{a}}{r}\right)^{1+k} \left(1 - \frac{\tilde{a}^2}{r^2}\right)^{\frac{1-k}{2}} n_x(\tilde{a}) d\tilde{a} \, \cos 2\theta \tag{29}$$

and for tangential displacements perpendicular to it the following formula holds

$$\bar{u}_y(r,\theta) = \frac{\pi(H-K)c_0^{\,k}}{(1+k)E_0} \int_0^{\mathrm{Min}\{r,a\}} \left(\frac{\tilde{a}}{r}\right)^{1+k} \left(1 - \frac{\tilde{a}^2}{r^2}\right)^{\frac{1-k}{2}} n_x(\tilde{a}) d\tilde{a} \, \sin 2\theta . \tag{30}$$

Equations (29) and (30) reveal some special features. First of all, it should be mentioned that the first two integrals in Eq. (29) indicate the axisymmetric part in the magnitude of the tangential displacements in the direction of loading, while the third one represents a harmonic alternating part with twice the polar angle according to $\cos 2\theta$. The tangential displacement perpendicular to loading, Eq. (30), has only a harmonic alternating component which coincides with that of the tangential displacement in $x$-direction apart from a phase shift by $-\pi/2$. A shortened version of Eqs. (29) and (30) reads

$$\begin{aligned} \bar{u}_x(r,\theta) &= g_1(r) + g_2(r)\cos 2\theta \\ \bar{u}_y(r,\theta) &= g_2(r)\sin 2\theta \end{aligned} , \tag{31}$$

where $g_1$ denotes the axisymmetric part in the magnitude of the tangential displacements in-$x$-direction and $g_2$ the amplitude of the harmonic parts. The latter is zero if the material parameters $H$ and $K$ are equal, which implies a very strong limitation[1]. It should be stressed that the dependence of the tangential displacement components on the polar angle according to Eq. (31) is universal, i.e., it applies to any unidirectional tangential stress distribution with axisymmetric magnitude. While the derivation of this result by means of the superposition principle of Mossakovskii and Jäger is trivial, its derivation in other ways, e.g., by integration of Green's functions for point loading (see Sect. 2.1) or by using Hankel transformations, is much more complicated. Eq. (31) is of particular importance for the tangential contact of two axisymmetric power-law graded elastic bodies in the state of gross slip, which will be discussed in more detail below.

Now all that remains is to determine the unknown function $n_x(x)$ from the given boundary conditions. Since we have assumed unidirectional tangential stresses (axisymmetric in magnitude) prescribed over a circular area of radius $r$ on a power-law graded elastic half-space

$$\tau_{zx}(r) = \begin{cases} -q_x(r) & \text{for } r \le a \\ 0 & \text{for } r > a \end{cases}, \tag{32}$$

the unknown function must be calculated from Eq. (27) by applying the inverse generalized Abel transformation

$$n_x(\tilde{a}) = -\frac{2\cos\left(\frac{k\pi}{2}\right)}{\pi} \frac{d}{d\tilde{a}} \int_{\tilde{a}}^{a} \frac{r q_x(r)}{\left(r^2 - \tilde{a}^2\right)^{\frac{1+k}{2}}} dr \quad \text{for } 0 < \tilde{a} < a . \tag{33}$$

### 4.3. Application example: Sliding of a sphere on a half-space with a plane surface

As a simple application example, consider a power-law graded elastic sphere stationary sliding in $x$-direction on a half-space of equal material having a plane surface. To a first approximation, the shape of the sphere can be assumed to be parabolic. As we will see later, in this case the tangential stress distribution in the contact area is as follows:

$$q_x(r) = q_0 \left(1 - \frac{r^2}{a^2}\right)^{\frac{1+k}{2}} \quad \text{for } 0 \le r \le a . \tag{34}$$

In the following, the universal calculation method derived in the last section will now be applied to determine the effects on the half-space surface. Substituting Eq. (34) into Eq. (33) followed by a short calculation leads to the linear function

$$n_x(\tilde{a}) = \frac{(1+k)q_0}{a^{1+k}} \tilde{a} . \tag{35}$$

Application of Eq. (28) then gives the resultant force of the tangential stress distribution

$$F_x = \frac{2\pi q_0 a^2}{3+k} . \tag{36}$$

Performing the integrals in Eqs. (29) and (30) taking into account the result (35) yields the following tangential displacement components within the loaded area after some simplifications:

---

$$\bar{u}_x\left(r,\theta\right)=\frac{\pi^2\left(1+k\right)q_0c_0^k}{8\cos\left(\frac{k\pi}{2}\right)a^{1+k}E_0}\left[\left(H+K\right)\left(2a^2-\left(1+k\right)r^2\right)+\left(H-K\right)\left(1-k\right)\frac{r^2}{2}\cos 2\theta\right]\quad\text{for }r\leq a$$

$$\bar{u}_y\left(r,\theta\right)=\frac{\pi^2\left(1-k^2\right)q_0c_0^k}{16\cos\left(\frac{k\pi}{2}\right)a^{1+k}E_0}\left(H-K\right)r^2\sin 2\theta\qquad\qquad\qquad\text{for }r\leq a$$

(37)

Outside of the loaded area, i.e. for $r>a$, the following displacement components are caused:

$$\bar{u}_x\left(r,\theta\right)=\frac{\pi q_0c_0^k a^{1-k}}{\left(3+k\right)E_0}\left(\frac{a}{r}\right)^{1+k}\left[\left(H+K\right){}_2\mathrm{F}_1\left(\frac{1+k}{2},\frac{1+k}{2};\frac{5+k}{2};\frac{a^2}{r^2}\right)+\left(H-K\right){}_2\mathrm{F}_1\left(\frac{k-1}{2},\frac{3+k}{2};\frac{5+k}{2};\frac{a^2}{r^2}\right)\cos 2\theta\right]$$

$$\bar{u}_y\left(r,\theta\right)=\frac{\pi q_0c_0^k a^{1-k}}{\left(3+k\right)E_0}\left(\frac{a}{r}\right)^{1+k}\left(H-K\right){}_2\mathrm{F}_1\left(\frac{k-1}{2},\frac{3+k}{2};\frac{5+k}{2};\frac{a^2}{r^2}\right)\sin 2\theta$$

(38)

For the limiting case $k=0$, the material coefficients simplify according to Eq. (4). If, in addition, the following identities are considered,

$$_2\mathrm{F}_1\left(\frac{1}{2},\frac{1}{2};\frac{5}{2};\frac{a^2}{r^2}\right)=\frac{3}{4}\frac{r^2}{a^2}\sqrt{1-\frac{a^2}{r^2}}+\frac{3}{4}\frac{r^3}{a^3}\left(2\frac{a^2}{r^2}-1\right)\arcsin\left(\frac{a}{r}\right)$$

$$_2\mathrm{F}_1\left(\frac{-1}{2},\frac{3}{2};\frac{5}{2};\frac{a^2}{r^2}\right)=\frac{3}{8}\left(2-\frac{r^2}{a^2}\right)\sqrt{1-\frac{a^2}{r^2}}+\frac{3}{8}\frac{r^3}{a^3}\arcsin\left(\frac{a}{r}\right)$$

(39)

Eqs. (37) and (38) agree exactly with Eqs. (3.90) - (3.92) given by Johnson [29] for the homogeneous elastic half-space.[2]

## 5. Tangential contact of two power-law graded elastic bodies

### 5.1. Coupling effects of two power-law graded elastic half-spaces in contact

Let us now consider two power-law graded elastic bodies with the same exponent of elastic inhomogeneity $k$ and equal characteristic depths $c_0$, but with different Poisson's ratios $\nu_1$, $\nu_2$ and elastic parameters $E_{01}$, $E_{02}$. Both bodies should be assumed as half-spaces. Due to external forces, they make contact over an area $A$ transmitting the pressure $p$ and distributed tangential stresses $q_x$ and $q_y$. By applying the principle of superposition, taking into account the linear equations (2), the relative elastic surface displacements, defined by $\bar{\mathbf{u}}^{(r)}:=\bar{\mathbf{u}}^{(1)}-\bar{\mathbf{u}}^{(2)}$, can be calculated

$$\bar{u}_x^{(r)}=c_0^k\int_A\left[\left(\frac{K_1}{E_{01}}+\frac{K_2}{E_{02}}\right)\frac{q_x}{s^{1+k}}+\left(\frac{H_1-K_1}{E_{01}}+\frac{H_2-K_2}{E_{02}}\right)\left(\frac{\left(x-\tilde{x}\right)^2}{s^{3+k}}q_x+\frac{\left(x-\tilde{x}\right)\left(y-\tilde{y}\right)}{s^{3+k}}q_y\right)-\left(\frac{L_1}{E_{01}}-\frac{L_2}{E_{02}}\right)\frac{x-\tilde{x}}{s^{2+k}}p\right]dA$$

$$\bar{u}_y^{(r)}=c_0^k\int_A\left[\left(\frac{H_1-K_1}{E_{01}}+\frac{H_2-K_2}{E_{02}}\right)\left(\frac{\left(x-\tilde{x}\right)\left(y-\tilde{y}\right)}{s^{3+k}}q_x+\frac{\left(y-\tilde{y}\right)^2}{s^{3+k}}q_y\right)+\left(\frac{K_1}{E_{01}}+\frac{K_2}{E_{02}}\right)\frac{q_y}{s^{1+k}}-\left(\frac{L_1}{E_{01}}-\frac{L_2}{E_{02}}\right)\frac{y-\tilde{y}}{s^{2+k}}p\right]dA$$ (40).

$$\bar{u}_z^{(r)}=c_0^k\int_A\left[\left(\frac{L_1}{E_{01}}-\frac{L_2}{E_{02}}\right)\left(\frac{x-\tilde{x}}{s^{2+k}}q_x+\frac{y-\tilde{y}}{s^{2+k}}q_y\right)+\left(\frac{B_1}{E_{01}}+\frac{B_2}{E_{02}}\right)\frac{p}{s^{1+k}}\right]dA$$

From Eqs. (40) it is evident that in general there is a coupling between normal and tangential effects, i.e., normal pressure produces tangential displacements and tangential stresses cause normal displacements. A complete decoupling of the normal and tangential contact problem and thus a substantial simplification of the governing equations is possible if holds [18]

$$\frac{L_1}{E_{01}} = \frac{L_2}{E_{02}} . \tag{41}$$

---

[2] Note that there is a typo in Eqs. (3.92a) and (3.92b). $x$ and $y$ must be replaced by $x/r$ and $y/r$ respectively.

Condition (41) is satisfied in the following three special cases [19] in particular:

1. The power-law graded elastic materials of both bodies are equal: $E_{01} = E_{02} =: E_0$ and $\nu_1 = \nu_2 =: \nu$

2. One body is rigid and the other has a Poisson's ratio equal to the Holl ratio: $E_{0i} \to \infty$ and $\nu_j = 1/(2+k)$ with $i \neq j$

3. The Poisson's ratios of both materials are given by the Holl ratio: $\nu_1 = \nu_2 = 1/(2+k)$

It is worth noting that in the two latter cases complete decoupling is only possible for $k \geq 0$, since negative exponents $k$ would lead to Poisson's ratios greater than 0.5 and thus leave the physically permissible range from -1 to 0.5. Moreover, Eqs. (40) show that the coupling effect is quantified by two different (mismatch) parameters. While the influence of tangential traction on normal pressure is specified by

$$\beta_{D1} := \left( \frac{L_1}{E_{01}} - \frac{L_2}{E_{02}} \right) \bigg/ \left( \frac{B_1}{E_{01}} + \frac{B_2}{E_{02}} \right) , \tag{42}$$

the influence of the pressure on the tangential displacements is expressed by the parameter

$$\beta_{D2} := \left( \frac{L_1}{E_{01}} - \frac{L_2}{E_{02}} \right) \bigg/ \left( \frac{K_1}{E_{01}} + \frac{K_2}{E_{02}} \right) . \tag{43}$$

Both parameters differ only by the denominator and coincide with Dundurs' second constant in the special case of elastically homogeneous materials. Since $B > K$ for $k < 0$ and $B < K$ for $k > 0$, the effect of tangential traction on normal pressure is smaller for negative and greater for positive exponents of elastic inhomogeneity compared to the reverse coupling effect. Even if we take into account that the tangential stresses are limited by the product of the friction coefficient times the pressure and a very small friction coefficient is assumed, it is necessary to carefully check in which cases the influence of the tangential stresses on the pressure can be neglected, but not vice versa. Therefore, the applicability of Goodman's approximation [30] to power-law graded materials depends on the exponent of elastic inhomogeneity and requires a separate investigation. However, this is beyond the scope of the present work. In the following, two power-law graded elastically similar materials according to Eq. (41) are assumed to ensure complete decoupling. Based on this assumption, Eqs. (40) simplify to

$$\bar{u}_x^{(r)} := \bar{u}_x^{(1)} - \bar{u}_x^{(2)} = c_0^k \int_A \left[ \left( \frac{K_1}{E_{01}} + \frac{K_2}{E_{02}} \right) \frac{q_x}{s^{1+k}} + \left( \frac{H_1 - K_1}{E_{01}} + \frac{H_2 - K_2}{E_{02}} \right) \left( \frac{(x-\tilde{x})^2}{s^{3+k}} q_x + \frac{(x-\tilde{x})(y-\tilde{y})}{s^{3+k}} q_y \right) \right] dA , \tag{44}$$

$$\bar{u}_y^{(r)} := \bar{u}_y^{(1)} - \bar{u}_y^{(2)} = c_0^k \int_A \left[ \left( \frac{H_1 - K_1}{E_{01}} + \frac{H_2 - K_2}{E_{02}} \right) \left( \frac{(x-\tilde{x})(y-\tilde{y})}{s^{3+k}} q_x + \frac{(y-\tilde{y})^2}{s^{3+k}} q_y \right) + \left( \frac{K_1}{E_{01}} + \frac{K_2}{E_{02}} \right) \frac{q_y}{s^{1+k}} \right] dA , \tag{45}$$

$$\bar{u}_z^{(r)} := \bar{u}_z^{(1)} - \bar{u}_z^{(2)} = c_0^k \left( \frac{B_1}{E_{01}} + \frac{B_2}{E_{02}} \right) \int_A \frac{p}{s^{1+k}} dA . \tag{46}$$

From Eqs. (44) and (45) it is evident that, despite the assumption of power-law graded elastically similar materials, there is still a second coupling between the components of the tangential stresses themselves. Complete decoupling is only given under the assumption

$$\frac{H_1 - K_1}{E_{01}} + \frac{H_2 - K_2}{E_{02}} = 0 \,. \tag{47}$$

To show that this is a very strict assumption, let us consider equal materials. In this case, the requirement (47) reduces to $H - K = 0$. Fig. 2 illustrates the dependence between Poisson's ratio and exponent of elastic inhomogeneity that ensures complete decoupling. The special case of elastically homogeneous materials is highlighted separately and requires the absence of any Poisson's effects $(\nu = 0)$.

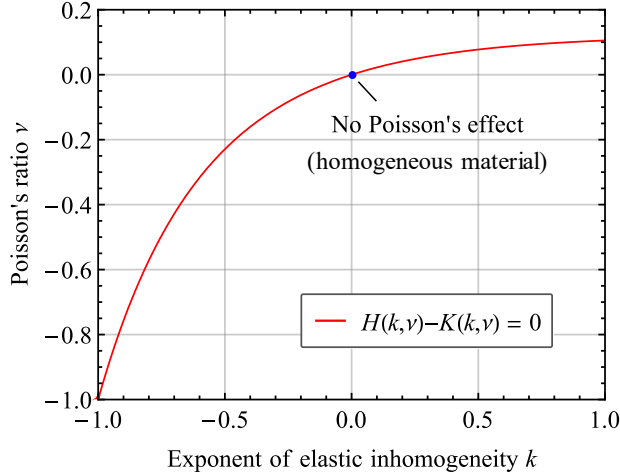

Fig. 2. Poisson's ratio as a function of the exponent of elastic inhomogeneity ensuring complete tangential decoupling

Taking into account the decoupling condition (47), Eqs. (44) and (45) simplify to

$$\overline{u}_j^{(r)}(x, y) = c_0^k \left( \frac{K_1}{E_{01}} + \frac{K_2}{E_{02}} \right) \int_A \frac{q_j(\tilde{x}, \tilde{y})}{s^{1+k}} dA \quad \text{with } j = x, y \,. \tag{48}$$

Note that Eq. (48) has the same form as Eq. (46) for normal contact, except for the different material parameters.

## 5.2. Partial slip contact between two solids made of power-law graded elastically similar materials

Let us now consider two solids of power-law graded elastically similar materials which are initially pressed into contact by a normal force $F_z$ forming a contact area $A$. Subsequently a monotonically increasing tangential force $F_x$ is applied, whose magnitude is not sufficient to cause full slip. A pure relative translation is assumed which requires non-rotating solids. The boundary condition for the normal contact reads

$$\overline{u}_z^{(r)}(x, y) = \delta_z - f(x, y) \quad \text{for } (x, y) \in A(F_z), \tag{49}$$

where $\delta_z$ denotes the normal contact approach and $f(x, y)$ the initial gap function. In the absence of adhesion, it is obvious that the pressure must be greater than zero everywhere within the contact area and zero outside. Moreover, the inequality of non-penetration must hold outside the contact area. Due to the applied monotonically increasing tangential force, the contact area is composed of (not a priori known) regions of stick and slip. Inside the stick region $A_{\text{stick}}$ the relative elastic tangential displacements must be constant and equal to the imposed rigid body translation[3]

---

[3] Tangential displacement of points within the two bodies distant from the contact interface caused by the applied tangential force.

$$\overline{u}_x^{(r)}(x,y) = \delta_x, \quad \overline{u}_y^{(r)}(x,y) = \delta_y \quad \text{for } (x,y) \in A_{\text{stick}} \subset A \,. \tag{50}$$

In addition, according to Amontons-Coulomb's law, within the stick zone the magnitude of the tangential tractions cannot exceed its limiting frictional value

$$|\boldsymbol{q}(x,y)| \leq \mu p(x,y) \quad \text{for } (x,y) \in A_{\text{stick}} \,, \tag{51}$$

where $\boldsymbol{q}(x,y) := q_x(x,y)\boldsymbol{e}_x + q_y(x,y)\boldsymbol{e}_y$. Within the slip region, however, the magnitude of the tangential tractions takes its limiting value

$$|\boldsymbol{q}(x,y)| = \mu p(x,y) \quad \text{for } (x,y) \in A_{\text{slip}} \,, \tag{52}$$

where $A_{\text{slip}} = A \setminus A_{\text{stick}}$. Furthermore, the slip $\boldsymbol{s}(x,y)$ must be directed opposed to the resultant tangential traction:

$$\frac{\boldsymbol{q}(x,y)}{|\boldsymbol{q}(x,y)|} = -\frac{\boldsymbol{s}(x,y)}{|\boldsymbol{s}(x,y)|} \quad \text{with} \quad \boldsymbol{s}(x,y) := \left(\overline{u}_x^{(r)}(x,y) - \delta_x\right)\boldsymbol{e}_x + \left(\overline{u}_y^{(r)}(x,y) - \delta_y\right)\boldsymbol{e}_y \,. \tag{53}$$

### 5.2.1. Partial slip under complete decoupling condition

Now let us suppose that, in addition to power-law graded elastically similar materials, the strict condition (47) is satisfied, so that complete decoupling exists and Eqs. (46) and (48) hold. For simplicity, it is assumed that the tangential stresses act only in the direction of the tangential displacement along the $x$-axis. Since $q_y$ is assumed to be zero, it follows from Eq. (48), putting $j = y$, that $u_y$ is zero as well. Substitution of Eq. (46) into the normal part of the boundary condition (48) yields

$$\delta_z - f(x,y) = c_0^k \left(\frac{B_1}{E_{01}} + \frac{B_2}{E_{02}}\right) \int_A \frac{p}{s^{1+k}} dA \quad \text{for } (x,y) \in A(F_z) \,, \tag{54}$$

from which the pressure distribution can be determined (in general numerically). Following the work of Ciavarella [23], the tangential stress distribution is assumed to consist of two parts, a full sliding one acting over the entire contact area and a correction part in the stick region

$$q_x(x,y) = \begin{cases} \mu p(x,y) - q_x^*(x,y) & \text{for } (x,y) \in A_{\text{stick}} \\ \mu p(x,y) & \text{for } (x,y) \in A_{\text{slip}} \end{cases} \,. \tag{55}$$

Considering this approach in Eq. (48), the boundary condition (50) gives

$$\delta_x = c_0^k \left(\frac{K_1}{E_{01}} + \frac{K_2}{E_{02}}\right) \left(\mu \int_A \frac{p(\tilde{x},\tilde{y})}{s^{1+k}} dA - \int_{A_{\text{stick}}} \frac{q_x^*(\tilde{x},\tilde{y})}{s^{1+k}} dA\right) \quad \text{for } (x,y) \in A_{\text{stick}} \,. \tag{56}$$

The first integral on the left side can be replaced by using the normal part of the boundary condition according to

Eq. (54) and simple rearranging results in

$$\delta_z - \frac{\delta_x}{\mu}\left(\frac{B_1}{E_{01}} + \frac{B_2}{E_{02}}\right)\bigg/\left(\frac{K_1}{E_{01}} + \frac{K_2}{E_{02}}\right) - f(x,y) = c_0^k\left(\frac{B_1}{E_{01}} + \frac{B_2}{E_{02}}\right)\int_{A_{\text{stick}}} \frac{q_x^*(\tilde{x},\tilde{y})/\mu}{s^{1+k}}\,dA \quad \text{for } (x,y) \in A_{\text{stick}}. \qquad (57)$$

Eq. (57) can be interpreted as being of the same kind as the original normal contact problem expressed by Eq. (54). The only differences are a smaller normal contact approach $\delta_z^*$ and a corresponding reduced contact area formed by the stick area. The correction term in the ansatz of the tangential stress distribution divided by the coefficient of friction is thus equal to the pressure distribution $p^*$ corresponding to a lower normal force $F_z^*$ which results in a reduced contact area identical with the stick area of the original problem

$$q_x^*(x,y)/\mu = p^*(x,y) \coloneqq p(x,y, A_{\text{stick}}). \qquad (58)$$

Inserting Eq. (58) into Eq. (55) yields

$$q_x(x,y) = \mu\big[p(x,y) - p(x,y,A_{\text{stick}})\big] \qquad (59)$$

and by means of subsequent simple integration we obtain

$$F_x = \mu\big(F_z - F_z^*\big) = \mu\big[F_z(A) - F_z(A_{\text{stick}})\big]. \qquad (60)$$

The corrective normal contact approach can be identified from Eq. (57). Hence, analogous to the tangential stress distribution and the tangential force, the relative tangential approach can be represented by the superposition of two normal contact approaches

$$\delta_x = \mu\alpha\big(\delta_z - \delta_z^*\big) = \mu\alpha\big[\delta_z(A) - \delta_z(A_{\text{stick}})\big] \quad \text{with } \alpha \coloneqq \left(\frac{K_1}{E_{01}} + \frac{K_2}{E_{02}}\right)\bigg/\left(\frac{B_1}{E_{01}} + \frac{B_2}{E_{02}}\right). \qquad (61)$$

Note that for elastically homogenous materials $\alpha = 1$ holds and Eqs. (59) to (61) agree with Ciavarella's results [23] for tangential loading of general three-dimensional contacts under negligible Poisson's ratio. The same equations are valid for partial slip contacts of half-planes (including multiple contact areas), which was proved by Jäger [31] and Ciavarella [32]. Furthermore, Jäger [25] and Ciavarella [33] extended the superposition principle for application to tangential contacts of elastically similar, axisymmetric bodies, taking tangential coupling into account. The only difference to Eqs. (59) to (61) is the value of $\alpha$, which is different from 1 and must be replaced by the ratio between the tangential and normal compliance. Based on well-founded arguments, Putignano et al. [26] even proposed to apply this extended version of the Ciavarella-Jäger theorem to the contact between two nominally flat but rough half-spaces under tangential coupling (i.e., for nonzero Poisson's ratio) as well. The required solution of the corresponding normal contact problem can be obtained, for example, by using the boundary element method. Such an approach was followed by Popov et al. [34]. It is worth noting that this procedure is completely compatible for application to power-law graded materials.

### 5.2.2. Partial slip contact between axisymmetric bodies in the presence of tangential coupling

In the following the partial slip contact between two axisymmetric solids is considered. In contrast to the last section, tangential coupling should be allowed. However, due to power-law graded elastically similar materials the normal contact is uncoupled so that the pressure distribution, normal contact approach and contact area caused by the normal force can be determined in advance by applying the theory proposed in Sect. 3.1. With the normal force kept constant, the subsequent application of a monotonically increasing tangential force results in microslip starting at the border of the contact area. It is worth repeating that a pure translation is assumed, i.e., two points within the bodies far from the contact interface undergo an imposed rigid-body translation. Within the stick area the relative elastic surface displacement must be equal to this rigid-body translation. Use is made of the simplifying assumption of Cattaneo and Mindlin that the tangential tractions act everywhere parallel to the applied tangential force which points in $x$-direction. Due to symmetry the rigid body displacement points in $x$-direction as well. Hence, the mixed boundary conditions of the tangential contact problem are

$$\bar{u}_x^{(r)}(r) = \delta_x, \quad \bar{u}_y^{(r)}(r) = 0 \quad \text{for } 0 \le r \le c, \tag{62}$$

$$|q_x(r)| = \mu p(r) \quad \text{for } c \le r \le a, \tag{63}$$

where $c$ and $a$ denote the stick radius and contact radius, respectively. Needless to say, conditions (51) and (53) must also be met. To solve the mixed boundary value problem according to Eqs. (62) and (63), once again the procedure from Mossakovskii and Jäger can be applied. As introduced in Chapt. 4, the general solution can be thought as a superposition of differential rigid shift motions. The tangential stresses within the slip annulus are obtained according to Eq. (27)

$$q_x(r) = \int_r^a \frac{n_x(\tilde{a})}{\left(\tilde{a}^2 - r^2\right)^{\frac{1-k}{2}}} d\tilde{a} \quad \text{for } c \le r \le a. \tag{64}$$

The tangential stresses within the stick region, on the other hand, are obtained by means of superposition of differential rigid shift solutions corresponding to radii between $c$ and $a$

$$q_x(r) = \int_c^a \frac{n_x(\tilde{a})}{\left(\tilde{a}^2 - r^2\right)^{\frac{1-k}{2}}} d\tilde{a} \quad \text{for } 0 \le r \le c. \tag{65}$$

Since according to the boundary condition (62) the relative surface displacement within the stick region should be constant, no differential solutions corresponding to radii smaller than $c$ are needed. By summing up the relevant differentials of Eq. (22), the relative displacement in the region of the stick reads

$$\bar{u}_x^{(r)}(r) = \frac{\pi^2 c_0^k}{2\cos\left(\frac{k\pi}{2}\right)} \left( \frac{H_1 + K_1}{E_{01}} + \frac{H_2 + K_2}{E_{02}} \right) \int_c^a n_x(\tilde{a}) d\tilde{a} \quad \text{for } 0 \le r \le c. \tag{66}$$

Substituting Eqs. (12) and (64) into (63) yields

$$\int_r^a \frac{n_x(\tilde{a})}{\left(\tilde{a}^2 - r^2\right)^{\frac{1-k}{2}}} d\tilde{a} = \mu \int_r^a \frac{n_z(\tilde{a})}{\left(\tilde{a}^2 - r^2\right)^{\frac{1-k}{2}}} d\tilde{a} \quad \text{for } c \le r \le a, \tag{67}$$

which is satisfied, if

$$n_x(\tilde{a}) = \mu n_z(\tilde{a}). \tag{68}$$

Substitution of Eq. (68) into (65) and subsequent splitting of the integral into two parts as well as comparison with Eq. (12) results in

$$q_x(r) = \mu \left[ p(r,a) - p(r,c) \right] \quad \text{for } 0 \le r \le c, \tag{69}$$

where $p(r,c)$ denotes the pressure generated by a (fictitious) smaller normal force associated with a smaller contact radius equal to the stick radius. By extending the domain of definition of the second pressure distribution to the slip region (where it is defined to be zero), it is allowed to write

$$q_x(r) = \mu \left[ p(r,a) - p(r,c) \right] \quad \text{for } 0 \le r \le a. \tag{70}$$

Integration of the tangential stresses over the contact area simply provides

$$F_x = \mu \left[ F_z(a) - F_z(c) \right]. \tag{71}$$

Substitution of Eq. (68) into (66) and subsequent splitting of the integral into two parts as well as taking Eq. (18) into account yields

$$\delta_x = \mu\tilde{\alpha}\left[\delta_z(a) - \delta_z(c)\right] \quad \text{with} \quad \tilde{\alpha} := \frac{1}{2}\left(\frac{H_1+K_1}{E_{01}} + \frac{H_2+K_2}{E_{02}}\right)\bigg/\left(\frac{B_1}{E_{01}} + \frac{B_2}{E_{02}}\right), \tag{72}$$

where $\tilde{\alpha}$ indicates the ratio of the tangential to the normal contact compliance. It is worth noting that in the case of tangential decoupling according to Eq. (47), $\tilde{\alpha}$ coincides with the factor $\alpha$ from Eq. (61). Results (69) to (71) involve the applicability of the Ciavarella-Jäger theorem to power-law graded elastic materials and were first presented by Heß [18] and published in the framework of the method of dimensionality reduction [19].

It is well known that the Cattaneo-Mindlin theory and its extension by Ciavarella and Jäger (under condition of tangential coupling) provide only approximate solutions, since the direction of slip within the annulus is not parallel to the assumed tangential traction, except on the axes. The slip vector can be noted as follows

$$\boldsymbol{s}(r,\theta) = s_x(r,\theta)\boldsymbol{e}_x + s_y(r,\theta)\boldsymbol{e}_y = \left(\bar{u}_x^{(r)}(r,\theta) - \delta_x\right)\boldsymbol{e}_x + \bar{u}_y^{(r)}(r,\theta)\boldsymbol{e}_y. \tag{73}$$

With the help of the Mossakovskii-Jäger procedure, its components can be calculated in a simple way as well. For this purpose, it is only necessary to note that points within the slip annulus lie outside the differential rigid shift areas with radii $c \le \tilde{a} \le r$ and inside those with radii $r \le \tilde{a} \le a$. Taking into account the corresponding different proportions from Eqs. (22) and (23) when composing the solutions, the slip components are as follows

$$s_x(r,\theta) = \frac{\pi c_0^{\ k}}{1+k}\left(\frac{H_1+K_1}{E_{01}} + \frac{H_2+K_2}{E_{02}}\right)\int_c^r\left[\left(\frac{\tilde{a}}{r}\right)^{1+k}{}_2F_1\left(\frac{1+k}{2},\frac{1+k}{2};\frac{3+k}{2};\frac{\tilde{a}^2}{r^2}\right) - \frac{\pi(k+1)}{2\cos\left(\frac{k\pi}{2}\right)}\right]n_x(\tilde{a})d\tilde{a}$$
$$+ \frac{\pi c_0^{\ k}}{1+k}\left(\frac{H_1-K_1}{E_{01}} + \frac{H_2-K_2}{E_{02}}\right)\int_c^r\left(\frac{\tilde{a}}{r}\right)^{1+k}\left(1-\frac{\tilde{a}^2}{r^2}\right)^{\frac{1-k}{2}}n_x(\tilde{a})d\tilde{a}\ \cos 2\theta \tag{74}$$

$$s_y(r,\theta) = \frac{\pi c_0^{\ k}}{1+k}\left(\frac{H_1-K_1}{E_{01}} + \frac{H_2-K_2}{E_{02}}\right)\int_c^r\left(\frac{\tilde{a}}{r}\right)^{1+k}\left(1-\frac{\tilde{a}^2}{r^2}\right)^{\frac{1-k}{2}}n_x(\tilde{a})d\tilde{a}\ \sin 2\theta. \tag{75}$$

The function $n_x(\tilde{a})$ in the integrand is proportional to the function $n_z(\tilde{a})$ according to Eq. (68). The latter has to be determined from the pure normal contact problem according to Eq. (20) which requires only knowledge about the initial gap function. We would like to emphasize once again that the above equations apply to partial slip contacts of arbitrarily shaped axisymmetric solids with a compact contact area. The component of transverse slip (74) can be understood as a measure of the error in the assumption that the tangential stresses are everywhere directed parallel to the $x$-axis. The importance of knowing the relative slip component according to Eq. (74) should be explicitly stressed, since multiplying it by the tangential traction gives the dissipated energy per unit area. The latter is an essential quantity for predicting the location of surface damage and crack nucleation.

## 6. Application example: Partial slip and energy dissipation in the contact of a sphere and a half-space made of elastically similar power-law graded material

As an example, we consider a sphere of radius $R$ pressed into contact with the plane surface of a half-space by a normal force $F_z$ forming a contact area of radius $a$. Subsequently, a monotonically increasing tangential force $F_x$ is applied. Both solids are made of power-law graded elastically similar material so that condition (41) holds and thus the normal and tangential contact problem are decoupled.

### 6.1. Solutions of the normal contact

As mentioned above, the key to solving the normal and partial slip contact problem lies in the determination of the function $n_z(\tilde{a})$. This requires the initial gap function of the contacting bodies in the undeformed configuration, which in our example is given by $f(r) = r^2/(2R)$. Substituting the gap function into Eq. (20) and considering the normal compliance of both bodies yields

$$n_z(\tilde{a}) = \frac{\cos\left(\frac{k\pi}{2}\right)}{\pi^2 c_0{}^k} \left(\frac{B_1}{E_{01}} + \frac{B_2}{E_{02}}\right)^{-1} \frac{2\tilde{a}}{(1+k)R}. \tag{76}$$

Insertion of Eq. (76) into Eqs. (12), (14) and (18) leads after basic integration to the solutions of the normal contact problem

$$p(r,a) = \frac{\cos\left(\frac{k\pi}{2}\right)}{\pi^2 c_0{}^k} \left(\frac{B_1}{E_{01}} + \frac{B_2}{E_{02}}\right)^{-1} \frac{2}{(1+k)^2 R} \left(a^2 - r^2\right)^{\frac{1+k}{2}}, \tag{77}$$

$$F_z(a) = \frac{4\cos\left(\frac{k\pi}{2}\right)}{\pi(3+k)(1+k)^2 Rc_0{}^k} \left(\frac{B_1}{E_{01}} + \frac{B_2}{E_{02}}\right)^{-1} a^{3+k}, \tag{78}$$

$$\delta_z(a) = \frac{a^2}{(1+k)R}. \tag{79}$$

### 6.2. Solution of the partial slip tangential contact

In Sect. 5.2.2, the applicability of the Ciavarella-Jäger theorem for partial slip contacts of power-law graded elastic solids has been proved. Hence, according to Eqs. (70) to (72), the solution of the problem can be represented by the superposition of the solutions of two normal contact problems. Taking Eqs. (77) to (79) into account, the following results are obtained

$$q_x(r) = \mu \frac{2\cos\left(\frac{k\pi}{2}\right)}{\pi^2(1+k)^2 c_0{}^k R} \left(\frac{B_1}{E_{01}} + \frac{B_2}{E_{02}}\right)^{-1} \left[\left(a^2 - r^2\right)^{\frac{1+k}{2}} - \left(c^2 - r^2\right)^{\frac{1+k}{2}}\right], \tag{80}$$

$$F_x = \mu \frac{4\cos\left(\frac{k\pi}{2}\right)}{\pi(3+k)(1+k)^2 Rc_0{}^k} \left(\frac{B_1}{E_{01}} + \frac{B_2}{E_{02}}\right)^{-1} \left(a^{3+k} - c^{3+k}\right) = \mu F_z \left[1 - \left(\frac{c}{a}\right)^{3+k}\right], \tag{81}$$

$$\delta_x = \frac{\mu\tilde{a}}{(1+k)R} \left(a^2 - c^2\right) = \mu\tilde{a}\delta_z \left(1 - \frac{c^2}{a^2}\right). \tag{82}$$

The stick radius normalized by the contact radius as a function of the normalized tangential force is plotted in Fig. 3. The blue curves indicate a negative exponent of elastic inhomogeneity, the red curves a positive one. It is obvious that the exponent has partly considerable influence. For example, a tangential force of 0.9375 times the maximum tangential force where gross slip sets in results in a ratio of 0.25 for $k = -1$ and a ratio twice as large for $k = 1$.

Fig. 4 illustrates the dependence of the normalized tangential force on the normalized tangential displacement for different exponents of the elastic inhomogeneity. One and the same normal force was assumed in all cases, resulting in different contact radii. The tangential displacement was normalized to the tangential displacement that occurs for homogeneous material when the state of gross slip is reached, $\delta_{x,0}^{\max}$. For $k = -1$ a linear curve is obtained, which is immediately evident from Eqs. (81) and (82). Furthermore, it can be clearly seen that the state of gross slip is reached for negative $k$ at smaller tangential displacements and for positive $k$ at larger tangential displacements compared to

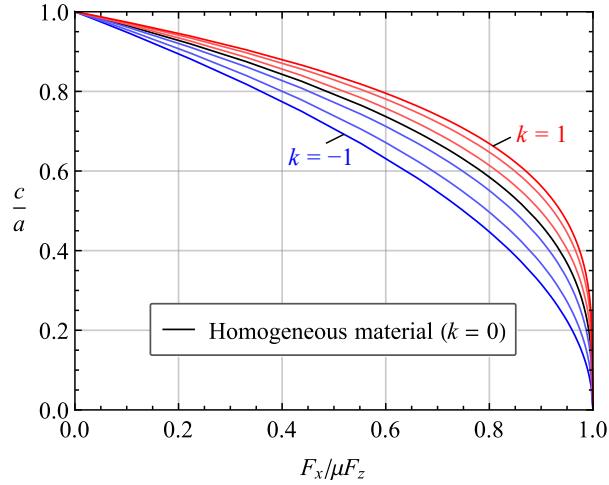

Fig. 3. Normalized representation of the stick radius as a function of
the tangential force for different exponents of elastic inhomogeneity homogeneous material. This effect, however, is strictly dependent on the chosen characteristic depth $c_0$, which here has been normalized to the Hertzian contact radius $a_0$. In Fig. 4, the characteristic depth was chosen equal to the Hertzian contact radius. For significantly smaller characteristic depths the effect reverses, i.e., for positive $k$ gross slip starts at a smaller displacement and for negative $k$ at a larger displacement. This is indicated by a comparison of Fig. 5a and Fig. 5b. Fig. 5 also shows the influence of the Poisson's ratio. At positive exponents of the elastic inhomogeneity the change of the Poisson's ratio leads to a noticeable spreading of the curves.

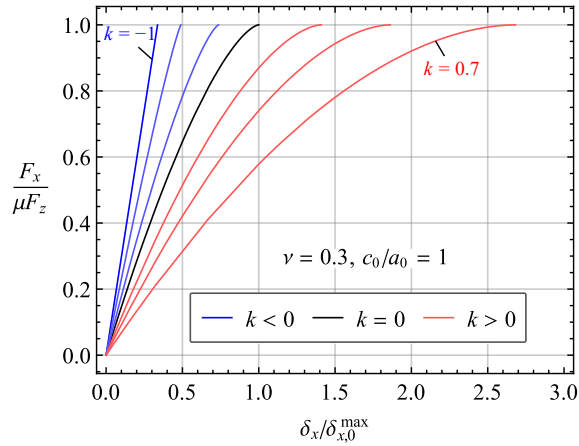

Fig. 4. Influence of the exponent of elastic inhomogeneity on the
normalized tangential force as a function of the normalized
tangential displacement

It should be noted that the Ciavarella-Jäger theorem can also be used to determine an approximation for the relative slip. This is possible if, in addition to the lateral slip component according to Eq. (75), the non-axially symmetric portion of the longitudinal slip in Eq. (74) is consequently neglected as well. Such an approximation is for instance used in the method of dimensionality reduction [35]. If one is interested in the mean loss energy under cyclic tangential loading, the non-axially symmetric term has no influence anyway. However, when studying crack initiation, especially the dissipated energy per unit area (surface density of dissipated energy) can become important.

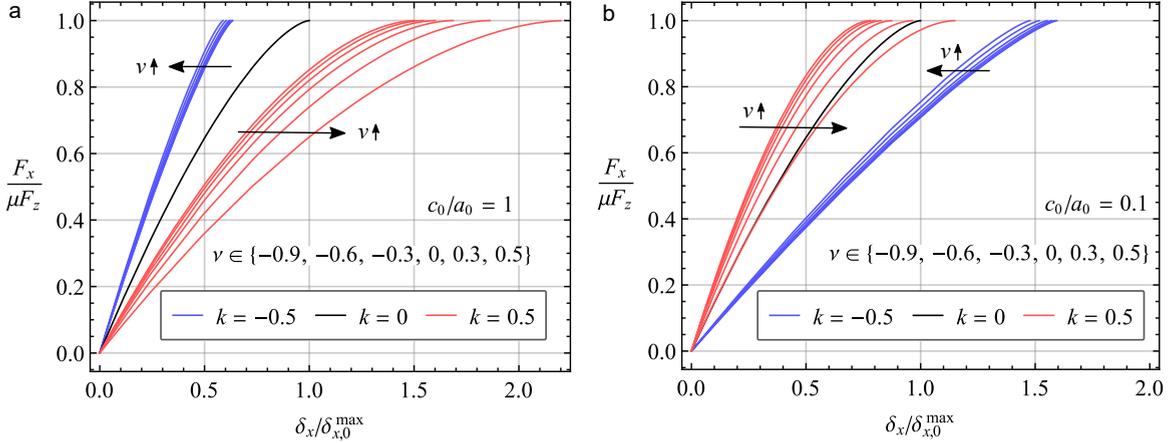

Fig. 5. Influence of the Poisson's ratio on the normalized tangential force as a function of the normalized tangential displacement for three representative exponents of elastic inhomogeneity and (a) a characteristic depth equal to the contact radius corresponding to homogeneous material, (b) a characteristic depth of one tenth of this radius.

### 6.3. Relative slip and density of dissipated energy

For the calculation of the relative slip components, once again merely the determined function $n_z(\tilde{a})$ according to Eq. (76) multiplied by the friction coefficient is required. After performing the integrals in Eqs. (74) and (75) we find

$$
s_x(r,\theta) = \mu \frac{F_z c_0{}^k}{2a^{3+k}} r^2 \left( \frac{H_1+K_1}{E_{01}} + \frac{H_2+K_2}{E_{02}} \right) \left[ \frac{\pi(1+k)(3+k)}{8\cos\left(\frac{k\pi}{2}\right)} \left( 2\frac{c^2}{r^2} - 1 - k \right) - \left(\frac{c}{r}\right)^{3+k} {}_2F_1\left( \frac{1+k}{2}, \frac{1+k}{2}; \frac{5+k}{2}; \frac{c^2}{r^2} \right) \right]
$$
$$
+ \mu \frac{F_z c_0{}^k}{2a^{3+k}} r^2 \left( \frac{H_1-K_1}{E_{01}} + \frac{H_2-K_2}{E_{02}} \right) \left[ \frac{\pi(1-k^2)(3+k)}{16\cos\left(\frac{k\pi}{2}\right)} - \left(\frac{c}{r}\right)^{3+k} {}_2F_1\left( \frac{k-1}{2}, \frac{3+k}{2}; \frac{5+k}{2}; \frac{c^2}{r^2} \right) \right] \cos 2\theta \tag{83}
$$

$$
s_y(r,\theta) = \mu \frac{F_z c_0{}^k}{2a^{3+k}} r^2 \left( \frac{H_1-K_1}{E_{01}} + \frac{H_2-K_2}{E_{02}} \right) \left[ \frac{\pi(1-k^2)(3+k)}{16\cos\left(\frac{k\pi}{2}\right)} - \left(\frac{c}{r}\right)^{3+k} {}_2F_1\left( \frac{k-1}{2}, \frac{3+k}{2}; \frac{5+k}{2}; \frac{c^2}{r^2} \right) \right] \sin 2\theta \tag{84}
$$

where we have used the expression (78) for the normal force to simplify the prefactors. Let us evaluate the equations for the special case where both bodies are made of equal elastically homogeneous material $\left( k=0, E_{0j}=E, \nu_j=\nu \right)$. Making use of the simplified material coefficients according to Eq. (4) as well as the mathematical equivalences in Eq. (39), the following relative slip components emerge

$$
s_x(r,\theta) = \frac{3\mu F_z}{32Ga^3} r^2\, 2(2-\nu) \left[ \left( 2\frac{c^2}{r^2} - 1 \right) \left( 1 - \frac{2}{\pi} \arcsin\left(\frac{c}{r}\right) \right) - \frac{2}{\pi} \frac{c}{r} \sqrt{1 - \frac{c^2}{r^2}} \right]
$$
$$
+ \frac{3\mu F_z}{32Ga^3} r^2 \nu \left[ 1 - \frac{2}{\pi} \arcsin\left(\frac{c}{r}\right) + \left( 1 - 2\frac{c^2}{r^2} \right) \frac{2}{\pi} \frac{c}{r} \sqrt{1 - \frac{c^2}{r^2}} \right] \cos 2\theta \tag{85}
$$

$$
s_y(r,\theta) = \frac{3\mu F_z}{32Ga^3} r^2 \nu \left[ 1 - \frac{2}{\pi} \arcsin\left(\frac{c}{r}\right) + \left( 1 - 2\frac{c^2}{r^2} \right) \frac{2}{\pi} \frac{c}{r} \sqrt{1 - \frac{c^2}{r^2}} \right] \sin 2\theta , \tag{86}
$$

where $G$ represents the shear modulus. The reader has surely noticed that these equations coincide exactly with the relative slip components derived by Johnson [36] for homogeneous material. Before proceeding with the study of the influence of elastic inhomogeneity, one interesting aspect for the special case of homogeneous material should be added. It is evident from Eq. (86) that except for a vanishing Poisson's ratio, there is a small lateral slip component, hence the condition that the slip must oppose the frictional traction is not precisely satisfied. However, this Mindlin-Johnson theory presupposes a priori that the direction of the tangential traction coincides with the direction of the applied tangential force. Using a numerical method, Munisamy et al. [21] showed that this assumption is not correct and that instead, especially in the partial slip regime, significant transverse tangential stresses occur, which become maximum on the lines $y = \pm x$. In Fig. 2 of their paper, the authors also present numerically calculated relative slip components for selected Poisson's ratios and polar angles. At this point, we would like to draw attention to the fact that the slip components according to Eqs. (85) and (86), which originate from the simplified theory, surprisingly agree quantitatively very well with the components resulting from the complete numerical computation of Munisamy et al. To the best of the authors' knowledge, this fact has not yet been pointed out and was probably first mentioned in a master's thesis [37] supervised by one of the authors (MH). To illustrate the good agreement, the slip components normalized to the maximum tangential displacement for $F_x = 0.8\mu F_z$ and at the onset of gross slip, i.e. for $F_x = \mu F_z$, are plotted in Fig. 6 for some of the specifications chosen by Munisamy et al.

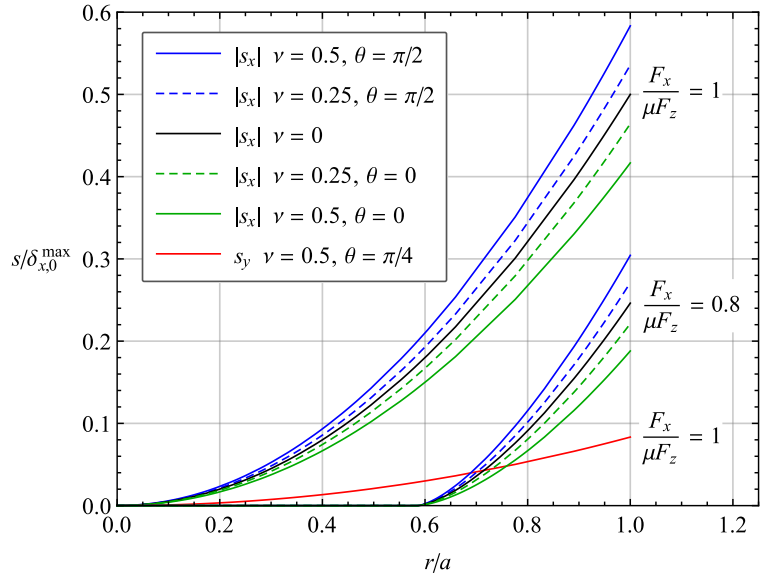

Fig. 6. Components of relative slip as a function of radial distance between parabolic solids of equal elastically homogeneous material: Influence of Poisson's ratio and polar angle

It is left to the reader to compare it with Fig. 2 from Munisamy et al. to confirm the close fit. Merely the $y$-component of relative slip seems to be a bit too high in the range of smaller radial distances. For incipient sliding and incompressible material Munisamy et al. state that the magnitude of the numerically calculated $y$-component at the edge of contact on lines $y = \pm x$ is about 15 percent of the magnitude of the $x$-component predicted by Eq. (85). A comparison with the analytically determined $y$-component, on the other hand, yields only a slightly higher value of 16.67 percent. For a common Poisson's ratio for steel of 0.3, an even smaller value of only 9 percent is obtained, which induced Johnson [36] to neglect the $y$-component and consequently the non-axisymmetric part of the $x$-component in Eq. (85) as well. Taking these assumptions into account, Eq. (85) provides the well-known approximation

$$s_x(r) \approx \frac{3\mu F_z}{32Ga^3} r^2 2(2-\nu)\left[\left(2\frac{c^2}{r^2}-1\right)\left(1-\frac{2}{\pi}\arcsin\left(\frac{c}{r}\right)\right)-\frac{2}{\pi}\frac{c}{r}\sqrt{1-\frac{c^2}{r^2}}\right].$$ (87)

Another outcome of the numerical calculations by Munisamy et al. is that the force-displacement characteristic for a full cycle of tangential loading deviates imperceptibly from the Mindlin approximation. Hence, the same holds true for the energy dissipated within a full cycle. However, this does not apply to the surface density of frictional energy dissipation defined by the scalar product

$$W_s(r,\theta) := 4\big|\mathbf{s}(r,\theta)\cdot\mathbf{q}(r,\theta)\big| \ . \tag{88}$$

For incipient sliding conditions and incompressible material Hills and Nowell [38] gave a contour plot of the energy dissipation over the contact area (see also Munisamy et al. [21]), which is distributed both symmetrically to the $x$-axis and $y$-axis. The numerical calculation predicts a significant difference between the maxima on the axes. While the value on the $y$-axis is slightly greater than 0.41, the maximum on the $x$-axis lies between 0.28 and 0.345. Clearly, the Mindlin approximation cannot represent this feature. Here, the slip and the tangential tractions are directed parallel to the applied external force and vary only with the radial and not with the angular position. For this reason, the energy density does not depend on the polar angle as well. However, if we use the non-approximated slip according to Eq. (85) instead of Mindlins approximation (87) to calculate the energy density, an excellent agreement with the numerical results is once again achieved. This is demonstrated by Fig. 7. The absolute maximum of 0.429 is located on the $y$-axis. It is 40 percent larger than the maximum on the $x$-axis which takes the value 0.306. Both maxima coincide well with the numerical results of Hills et al. The same holds for their normalized radial distance of 0.816 from the origin. Finally, we note that the absolute maximum for artificial materials characterized by negative Poisson's ratios is on the $x$-axis, which is immediately evident from the non-axisymmetric part of the slip given by Eq. (85).

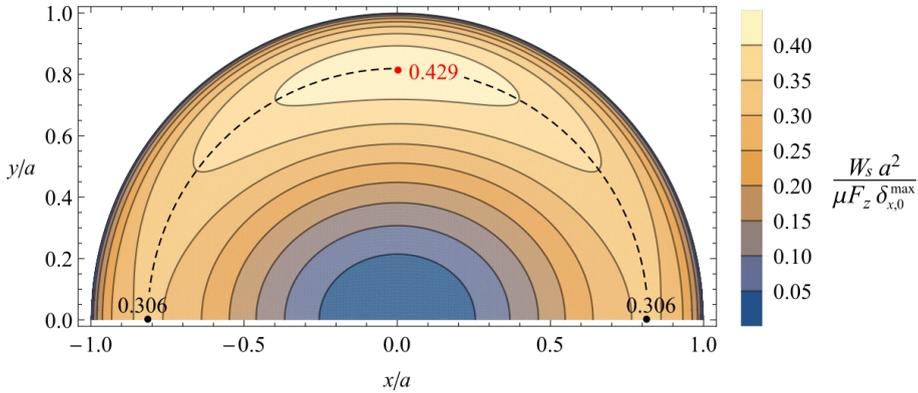

Fig. 7. Surface density of frictional energy dissipation in a tangential contact of parabolic shaped, elastically equal, and incompressible solids under incipient sliding conditions.

In the following, the above findings, namely that both the analytically calculated slip and the energy density for elastically homogeneous materials are in good agreement with the results of a full numerical calculation, are applied to power-law graded materials without any proof. We focus on the condition of incipient full sliding. In this state, the slip components for all exponents of elastic inhomogeneity increase quadratically with radial distance, which is evident from Eqs. (83) and (84) after setting $c = 0$

$$\frac{s_x(r,\theta)}{\delta_{x,0}^{\max}} = \frac{(1-\nu)}{2(2-\nu)}\frac{r^2}{Ba^2}\left[-H-K+\frac{(1-k)(H-K)}{2(1+k)}\cos 2\theta\right] \ , \tag{89}$$

$$\frac{s_y(r,\theta)}{\delta_{x,0}^{\max}} = \frac{(1-\nu)(1-k)(H-K)}{4(2-\nu)(1+k)Ba^2}r^2\sin 2\theta \ , \tag{90}$$

where we have normalized to the maximum tangential displacement for homogeneous material and assumed equal materials. Since it is assumed, in accordance with the Cattaneo-Mindlin theory, that the tangential tractions act parallel to the applied tangential force, only the $x$-component of the slip is needed to calculate the surface density of energy dissipation. From Eq. (89) it is easy to see that the magnitude of the slip is composed of an axisymmetric component and another one that varies harmonically in circumferential direction. Whereas the axisymmetric part is dominated by the ratio of the tangential to the normal contact compliance characterized by the parameter $\tilde{\alpha}$ in Eq. (72), the magnitude of the non-axisymmetric part is dictated by the tangential coupling parameter $H - K$. Due to its harmonically alternating character, the latter has no influence on the total energy dissipated within a full cycle, but it does have a decisive influence on the surface density of energy dissipation. For grading with a negative exponent $k = -0.5$ and two different Poisson's ratios this is illustrated in Fig. 8. The maxima are about a factor of 10 smaller than for homogeneous materials. Whereas for a negative Poisson's ratio of $\nu = -0.5$, which characterizes artificial materials, the absolute maximum lies on the $x$-axis (see Fig. 8a), for a positive Poisson's ratio it is located on the $y$-axis (see Fig. 8b). This tendency is observed irrespective of the exponent of elastic inhomogeneity (compare with Fig. 9). For incompressible material, $\nu = 0.5$, the maximum on the $y$-axis exceeds the maximum on the $x$-axis by more than 50 percent.

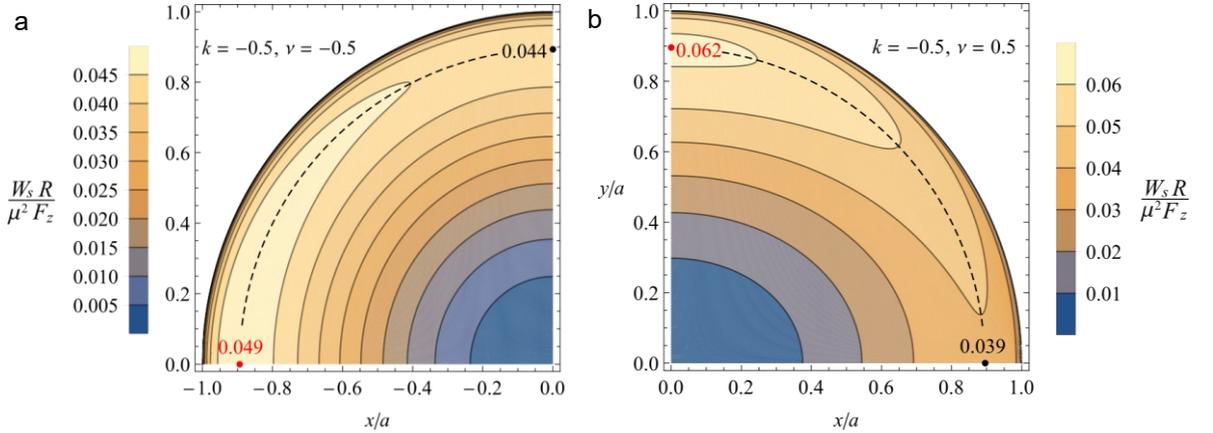

Fig. 8. Surface density of frictional energy dissipation in a tangential contact of parabolic shaped solids of equal power-law graded material characterized by a negative exponent of elastic inhomogeneity $k = -0.5$ under incipient sliding conditions for (a) a negative Poisson's ratio of $\nu = -0.5$ and (b) a positive Poisson's ratio of $\nu = 0.5$

Fig. 9 shows for comparison the dissipated energy density for a positive exponent of $k = 0.5$. For negative Poisson's ratios, the maxima on the axes are slightly greater than twice those for homogeneous material (see Fig. 9a). For incompressible material, the value is even four times higher (Fig. 9b). The maxima on the axes differ by about 20 percent. Furthermore, a comparison of Fig. 7 to Fig. 9 reveals that the radial distance of the maxima decreases with increasing exponent of elastic inhomogeneity. This is due to the stress distribution according to Eqs. (77) and (80), respectively. If we assume equal normal forces and equal contact radii, then the maximum becomes smaller with decreasing exponents and the curve becomes broader in the center region.

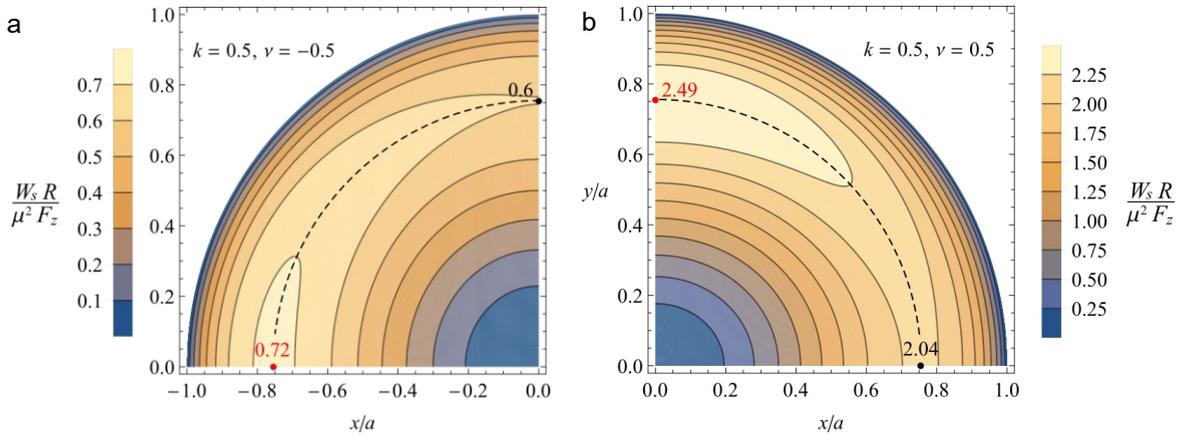

Fig. 9. Surface density of frictional energy dissipation in a tangential contact of parabolic shaped solids of equal power-law graded material characterized by a positive exponent of elastic inhomogeneity $k = 0.5$ under incipient sliding conditions for (a) a negative Poisson's ratio of $\nu = -0.5$ and (b) a positive Poisson's ratio of $\nu = 0.5$

## 7. Conclusions and outlook

In recent years, contact mechanics of local isotropically, power-law graded elastic solids has been widely advanced. Closed analytical solutions have been developed for axisymmetric normal contacts, taking adhesion into account as well. On this basis, important concepts and theories originally developed for contact problems of homogeneous elastic materials could be extended in a straightforward way to power-law graded materials [39] [40]. However, up to now there was a gap in the treatment of tangential contacts of power-law graded elastic solids, which has been closed by the general theory presented here. It allows a simple calculation of closed-form analytic solutions for all relevant quantities of tangential contacts between power-law graded elastically similar axisymmetric bodies under different boundary conditions and includes the (theoretically) exact determination of local quantities such as the tangential surface displacements or the relative slip components. The knowledge of the latter is important for the surface density of energy dissipation, which is one of the relevant factors for the analysis of crack initiation in fretting. Within the framework of an application example, an extended approach for the calculation of the local dissipated energy has been proposed, which, in the special case of elastically homogeneous material, yields results that agree very well with those reported in the literature obtained from a full numerical computation. Our approach makes use of the usual assumption that the tangential traction in the contact area act everywhere parallel to the applied tangential force, however, it accounts for the non-axisymmetric portion of the magnitude of relative slip in addition to the axisymmetric one. Whereas the axisymmetric part is dominated by the ratio of the tangential to the normal contact compliance, the magnitude of the non-axisymmetric part is dominated by the coupling parameter of tangential stresses themselves. Based on the new approach, the influence of both Poisson's ratio and elastic inhomogeneity on the surface density of energy dissipation within a contact of parabolically shaped power-law graded elastic solids under (cyclic) tangential loading up to incipient sliding was investigated. It is found that for a grading with a negative exponent $k = -0.5$ the maxima of energy density are about 10 times smaller than for homogeneous materials whereas for a grading with a positive exponent of $k = 0.5$ they are significant larger. For incompressible material, $\nu = 0.5$, and a negative exponent $k = -0.5$ the maximum on the $y$-axis exceeds the maximum on the $x$-axis by more than 50 percent. Hence, the Mindlin approximation cannot be used in this case since it neglects the non-axisymmetric part.

Another result concerns the position of the maxima. The radial distance of the maxima of energy density decreases with increasing exponent of the elastic inhomogeneity. However, the circumferential position depends decisively on the sign of the Poisson's ratio, independent of the sign of the elastic inhomogeneity. For artificial material characterized by a negative Poisson's ratio the absolute maximum lies on the $x$-axis, for a positive one it is located on the $y$-axis. It should be emphasized once again that all calculations refer to power-law graded elastically similar materials. The applicability of Goodman's approximation to power-law graded dissimilar materials needs a separate investigation since it depends on the exponent of elastic inhomogeneity. In principle, solving tangential contact problems between dissimilar power-law graded elastic solids requires numerical methods because the coupling effects impose incremental procedures for determining stick and slip regions. For this purpose, in a future work it is intended to extend the boundary element method for graded materials introduced by Li and Popov [41], taking into account the incremental numerical procedure proposed by Munisamy et al. [21].

An extension of the method including the adapted Ciavarella-Jäger theorem for axisymmetric as well as rough contacts, to transversely isotropic power-law graded elastic materials can be easily achieved. Therefore, only an adjustment of the material parameters is necessary, quite analogous to that for homogeneous materials [42] [43]. The general method presented here paves the way for extending various important works, originally developed for homogeneous materials, to power-law graded materials. For partial validation of the results, the limiting case $k = 0$ can be used since the solutions must coincide with those for elastically homogeneous solids.

## Author contributions

Markus Heß: Conceptualization, Formal analysis, Investigation, Methodology, Writing- original draft, Writing-review & editing. Qiang Li: Investigation, Software, Validation

## Acknowledgements

The authors would like to thank Kai Simbruner and Christian Jahnke for many helpful discussions in the context of their master's theses from 2017 and 2018, respectively.

## Appendix A. Derivation of solutions for a uniform tangential shift of the loaded circular area

For application of the Mossakovskii-Jäger procedure in Section 4.2 the solution for a uniform tangential shift of the loaded circular area is needed, which is deduced below. For this purpose, the uni-directional tangential traction

$$q_x(r) = \frac{q_0 a^{1-k}}{\left(a^2 - r^2\right)^{\frac{1-k}{2}}} \tag{91}$$

is applied to a circular region of radius $a$ on the power-law graded elastic half-space. The tangential surface displacements in the direction of loading as well as perpendicular to it are calculated by superposition of point force solutions introduced in Chapt. 2:

$$\bar{u}_x(x,y) = \frac{c_0^k}{E_0} \int_A \left( \frac{K}{s^{1+k}} q_x(\tilde{x}, \tilde{y}) + (H - K) \frac{(x - \tilde{x})^2}{s^{3+k}} q_x(\tilde{x}, \tilde{y}) \right) dA \quad, \tag{92}$$

$$\bar{u}_y(x,y) = \frac{c_0^k}{E_0} \int_A (H - K) \frac{(x - \tilde{x})(y - \tilde{y})}{s^{3+k}} q_x(\tilde{x}, \tilde{y}) dA \quad. \tag{93}$$

As usual (see [29]), distinction is made between a calculation of displacements inside and outside of the loading area.

Fig. 10 lists the integration variables used in both cases.

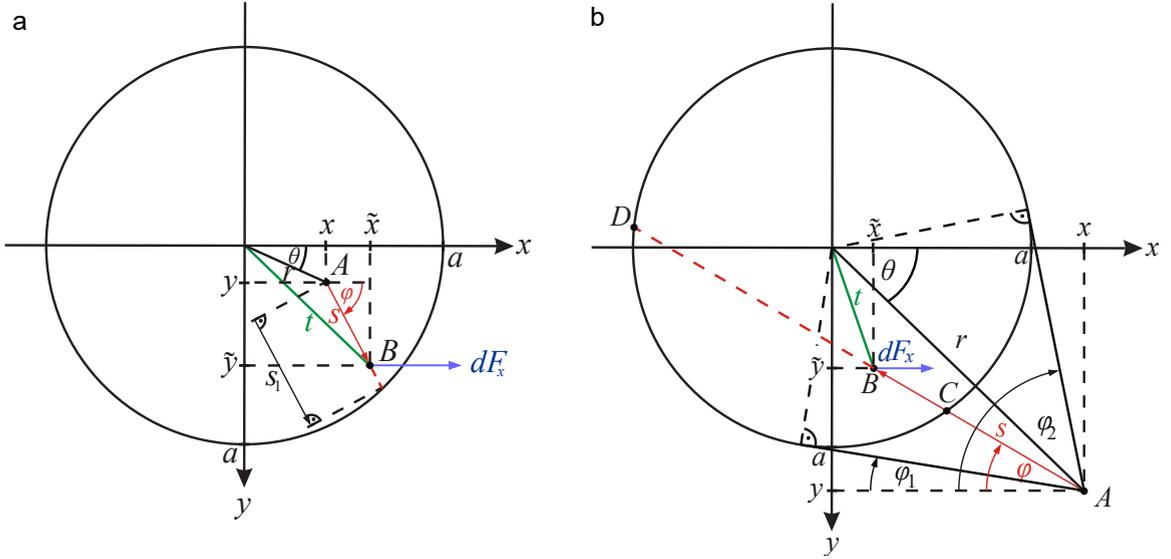

Fig. 10. Auxiliary diagrams to explain the integral determination of the tangential displacement components (a) inside and (b) outside the circular area which is loaded by uni-directional tangential tractions.

### A.1. Tangential surface displacements inside the loaded circular region

The circular area loaded by the unidirectional tangential traction is shown in Fig. 10a. The displacements of a point A within the area are sought. In point B, a load increment $dF_x = q_x(\tilde{x}, \tilde{y})dA$ of the entire distribution is highlighted exemplarily. Using a transformation to the coordinates $s$ and $\varphi$ shown in Fig. 10a, Eqs. (92) and (93) become

$$\bar{u}_x(r,\theta) = \frac{c_0^k}{E_0} \int_0^{2\pi} \left(K + (H-K)\cos^2\varphi\right) \int_0^{s_1(\varphi)} q_x\left(t(s,\varphi)\right)s^{-k}\,ds\,d\varphi \;, \tag{94}$$

$$\bar{u}_y(r,\theta) = (H-K)\frac{c_0^k}{E_0} \int_0^{2\pi} \sin\varphi\cos\varphi \int_0^{s_1(\varphi)} q_x\left(t(s,\varphi)\right)s^{-k}\,ds\,d\varphi \;. \tag{95}$$

Using the cosine theorem, the distance of the moving load point B from the origin is given by

$$t^2(s,\varphi) = r^2 + s^2 + 2rs\cos(\varphi-\theta) \;. \tag{96}$$

By introducing $\alpha^2 = a^2 - r^2$ and $\beta(\varphi) = r\cos(\varphi-\theta)$ as well as taking Eqs. (91) and (96) into account, Eqs. (94) and (95) can be written as follows

$$\bar{u}_x(r,\theta) = \frac{q_0 a^{1-k} c_0^k}{E_0} \int_0^{2\pi} \left(K + (H-K)\cos^2\varphi\right) \int_0^{s_1(\varphi)} \frac{s^{-k}}{\left(\sqrt{\alpha^2 - 2\beta s - s^2}\right)^{1-k}}\,ds\,d\varphi \;, \tag{97}$$

$$\bar{u}_y(r,\theta) = (H-K)\frac{q_0 a^{1-k} c_0^k}{E_0} \int_0^{2\pi} \sin\varphi\cos\varphi \int_0^{s_1(\varphi)} \frac{s^{-k}}{\left(\sqrt{\alpha^2 - 2\beta s - s^2}\right)^{1-k}}\,ds\,d\varphi \;, \tag{98}$$

where $s_1 = -\beta + \sqrt{\beta^2 + \alpha^2}$ is the positive solution of the condition $t(s_1, \varphi) \overset{!}{=} a$ (see Fig. 10a). The inner integral over $s$ can be substituted by $u := (s + \beta) / \sqrt{\alpha^2 + \beta^2}$ and then determined analytically

$$\int_0^{s_1(\varphi)} \frac{s^{-k}}{\left(\sqrt{\alpha^2 - 2\beta s - s^2}\right)^{1-k}} ds = \int_C^1 \frac{1}{\left(1 - u^2\right)^{\frac{1-k}{2}} (u - C)^k} du = \frac{\pi}{2\cos\left(\frac{k\pi}{2}\right)} - \frac{\Gamma\left(1 - \frac{k}{2}\right)\Gamma\left(\frac{1+k}{2}\right)}{\sqrt{\pi}} C\, {}_2F_1\left(\frac{1}{2}, \frac{1+k}{2}; \frac{3}{2}; C^2\right) \quad (99)$$

where $C(\varphi) := \beta(\varphi) / \sqrt{\alpha^2 + \beta(\varphi)^2}$. Substituting Eq. (99) into (97) and (98) and taking the symmetry relations $C(\varphi + \pi) = -C(\varphi)$, $C(\varphi + \pi)^2 = C(\varphi)^2$, $\cos^2(\varphi + \pi) = \cos^2(\varphi)$ and $\sin(\varphi + \pi)\cos(\varphi + \pi) = \sin(\varphi)\cos(\varphi)$ into account, the hypergeometric term will not contribute after integrating over $\varphi$. The remaining elementary integrals result in

$$\bar{u}_x(r, \theta) = \frac{\pi q_0 a^{1-k} c_0^k}{2\cos\left(\frac{k\pi}{2}\right) E_0} \int_0^{2\pi} \left(K + (H - K)\cos^2\varphi\right) d\varphi = \frac{\pi^2 q_0 a^{1-k} c_0^k}{2\cos\left(\frac{k\pi}{2}\right) E_0} (H + K) \quad , \tag{100}$$

$$\bar{u}_y(r, \theta) = (H - K)\frac{\pi q_0 a^{1-k} c_0^k}{2\cos\left(\frac{k\pi}{2}\right) E_0} \int_0^{2\pi} \sin\varphi\cos\varphi\, d\varphi = 0 \quad . \tag{101}$$

Thus, there is actually a rigid-body translation of the entire loading area in the direction of the applied tangential traction.

*A.2. Tangential surface displacements outside the loaded circular region*

For the calculation of the displacements of a point A outside the loading area, the integral equations (92) and (93) are used once again, but now considering the integration variables declared in Fig. 10b. This results in

$$\bar{u}_x(r, \theta) = \frac{c_0^k}{E_0} \int_{\varphi_1}^{\varphi_2} \left(K + (H - K)\cos^2\varphi\right) \int_{s_C}^{s_D} q_x\left(t(s, \varphi)\right) s^{-k}\, ds\, d\varphi \quad , \tag{102}$$

$$\bar{u}_y(r, \theta) = (H - K)\frac{c_0^k}{E_0} \int_{\varphi_1}^{\varphi_2} \sin\varphi\cos\varphi \int_{s_C}^{s_D} q_x\left(t(s, \varphi)\right) s^{-k}\, ds\, d\varphi \quad . \tag{103}$$

The distance of the moving load point B from the origin is now given by

$$t^2(s, \varphi) = r^2 + s^2 - 2rs\cos(\theta - \varphi) \quad . \tag{104}$$

The limits of the integral over $s$ are indicated by points C and D in Figure 10b and can be determined from the condition $t(s, \varphi) \overset{!}{=} a$. In this way we obtain $s_{D,C} = \beta \pm \sqrt{\beta^2 - \alpha^2}$ with $\alpha^2 = r^2 - a^2$ and $\beta(\varphi) = r\cos(\theta - \varphi)$ introduced. Furthermore, from simple geometry in Fig. 10b, $\varphi_{1,2} = \theta \mp \arcsin(a/r)$ hold. Considering Eq. (104) in (91) and substituting the result into Eqs. (102) and (103) provides

$$\bar{u}_x(r, \theta) = \frac{q_0 a^{1-k} c_0^k}{E_0} \int_{\varphi_1}^{\varphi_2} \left(K + (H - K)\cos^2\varphi\right) \int_{s_C}^{s_D} \frac{s^{-k}}{\left(\sqrt{-\alpha^2 + 2\beta s - s^2}\right)^{1-k}}\, ds\, d\varphi \quad , \tag{105}$$

$$\bar{u}_y(r, \theta) = (H - K)\frac{q_0 a^{1-k} c_0^k}{E_0} \int_{\varphi_1}^{\varphi_2} \sin\varphi\cos\varphi \int_{s_C}^{s_D} \frac{s^{-k}}{\left(\sqrt{-\alpha^2 + 2\beta s - s^2}\right)^{1-k}}\, ds\, d\varphi \quad . \tag{106}$$

The inner integral over $s$ can be substituted by $u := (s - \beta)/\sqrt{\beta^2 - \alpha^2}$ and then determined analytically

$$\int_{s_C}^{s_D} \frac{s^{-k}}{\left(\sqrt{-\alpha^2 + 2\beta s - s^2}\right)^{1-k}} ds = \int_{-1}^{1} \frac{1}{\left(1 - u^2\right)^{\frac{1-k}{2}} \left(u + C\right)^k} du = \frac{\sqrt{\pi}\,\Gamma\left(\frac{1+k}{2}\right)}{\Gamma\left(1 + \frac{k}{2}\right)} C^{-k} \, {}_2F_1\left(\frac{k}{2}, \frac{1+k}{2}; 1 + \frac{k}{2}; \frac{1}{C^2}\right), \quad (107)$$

where $C(\varphi) := \beta(\varphi)/\sqrt{\beta(\varphi)^2 - \alpha^2}$. Inserting Eq. (107) into (105) and (106) followed by substitution $\gamma := \theta - \varphi$ lead to

$$\bar{u}_x(r, \theta) = \frac{\sqrt{\pi}\,q_0 a^{1-k} c_0^k}{E_0} \frac{\Gamma\left(\frac{1+k}{2}\right)}{\Gamma\left(1 + \frac{k}{2}\right)} \int_{-\bar{\gamma}}^{\bar{\gamma}} \left[K + (H - K)\cos^2(\theta - \gamma)\right] C(\gamma)^{-k} \, {}_2F_1\left(\frac{k}{2}, \frac{1+k}{2}; 1 + \frac{k}{2}; \frac{1}{C(\gamma)^2}\right) d\gamma, \quad (108)$$

$$\bar{u}_y(r, \theta) = (H - K)\frac{\sqrt{\pi}\,q_0 a^{1-k} c_0^k}{E_0} \frac{\Gamma\left(\frac{1+k}{2}\right)}{\Gamma\left(1 + \frac{k}{2}\right)} \int_{-\bar{\gamma}}^{\bar{\gamma}} \sin(\theta - \gamma)\cos(\theta - \gamma) C(\gamma)^{-k} \, {}_2F_1\left(\frac{k}{2}, \frac{1+k}{2}; 1 + \frac{k}{2}; \frac{1}{C(\gamma)^2}\right) d\gamma, \quad (109)$$

where $C(\gamma) := C(\varphi(\gamma))$ and $\bar{\gamma} = \arcsin(a/r)$. Using the property $C(-\gamma) = C(\gamma)$ and addition theorems, the above equations can be simplified as follows

$$\begin{aligned}
\bar{u}_x(r, \theta) &= \frac{\sqrt{\pi}\,q_0 a^{1-k} c_0^k}{E_0} \frac{\Gamma\left(\frac{1+k}{2}\right)}{\Gamma\left(1 + \frac{k}{2}\right)} (H + K)\int_0^{\bar{\gamma}} C(\gamma)^{-k} \, {}_2F_1\left(\frac{k}{2}, \frac{1+k}{2}; 1 + \frac{k}{2}; \frac{1}{C(\gamma)^2}\right) d\gamma \\
&+ \frac{\sqrt{\pi}\,q_0 a^{1-k} c_0^k}{E_0} \frac{\Gamma\left(\frac{1+k}{2}\right)}{\Gamma\left(1 + \frac{k}{2}\right)} (H - K)\cos(2\theta)\int_0^{\bar{\gamma}} \cos(2\gamma) C(\gamma)^{-k} \, {}_2F_1\left(\frac{k}{2}, \frac{1+k}{2}; 1 + \frac{k}{2}; \frac{1}{C(\gamma)^2}\right) d\gamma
\end{aligned} \quad (110)$$

$$\bar{u}_y(r, \theta) = \frac{\sqrt{\pi}\,q_0 a^{1-k} c_0^k}{E_0} \frac{\Gamma\left(\frac{1+k}{2}\right)}{\Gamma\left(1 + \frac{k}{2}\right)} (H - K)\sin(2\theta)\int_0^{\bar{\gamma}} \cos(2\gamma) C(\gamma)^{-k} \, {}_2F_1\left(\frac{k}{2}, \frac{1+k}{2}; 1 + \frac{k}{2}; \frac{1}{C(\gamma)^2}\right) d\gamma. \quad (111)$$

Note that the integral in Eq. (111) is the same as the second one in (110). But first we focus on the calculation of the first integral $I_1$ in Eq. (110). For this purpose, the substitution $z = 1/C(\gamma)^2$ is applied

$$I_1 := \int_0^{\bar{\gamma}} C(\gamma)^{-k} \, {}_2F_1\left(\frac{k}{2}, \frac{1+k}{2}; 1 + \frac{k}{2}; \frac{1}{C(\gamma)^2}\right) d\gamma = -\frac{1}{2}\sqrt{1 - \frac{a^2}{r^2}} \int_{a^2/r^2}^0 \frac{{}_2F_1\left(\frac{k}{2}, \frac{1+k}{2}; 1 + \frac{k}{2}; z\right) z^{k/2}}{(1 - z)\sqrt{(a/r)^2 - z}} dz. \quad (112)$$

By focus on a single term of the power series of the hypergeometric function in the integrand, then performing the integral, and finally sum up, after elaborate calculation the following result is obtained

$$I_1 = \frac{\sqrt{\pi}\,\Gamma\left(1 + \frac{k}{2}\right)}{(k+1)\Gamma\left(\frac{1+k}{2}\right)} \left(\frac{a}{r}\right)^{k+1} \, {}_2F_1\left(\frac{1+k}{2}, \frac{1+k}{2}; \frac{3+k}{2}; \frac{a^2}{r^2}\right). \quad (113)$$

The second integral $I_2$ can be expressed by

$$I_2 := \int_0^{\bar{\gamma}} \frac{\cos(2\gamma)}{C(\gamma)^k} \, {}_2F_1\left(\frac{k}{2}, \frac{1+k}{2}; 1 + \frac{k}{2}; \frac{1}{C(\gamma)^2}\right) d\gamma = 2\int_0^{\bar{\gamma}} \frac{\cos^2\gamma}{C(\gamma)^k} \, {}_2F_1\left(\frac{k}{2}, \frac{1+k}{2}; 1 + \frac{k}{2}; \frac{1}{C(\gamma)^2}\right) d\gamma - I_1. \quad (114)$$

It is composed of the already known integral $I_1$ and another integral $I_3$, for its solution the substitution

$z = 1/C(\gamma)^2$ is applied once again:

$$I_3 := 2\int_0^{\overline{\gamma}} \frac{\cos^2\gamma}{C(\gamma)^k} {}_2F_1\left(\frac{k}{2}, \frac{1+k}{2}; 1+\frac{k}{2}; \frac{1}{C(\gamma)^2}\right) d\gamma = -\left(1 - \frac{a^2}{r^2}\right)^{3/2} \int_{a^2/r^2}^0 \frac{{}_2F_1\left(\frac{k}{2}, \frac{1+k}{2}; 1+\frac{k}{2}; z\right) z^{k/2}}{(1-z)^2 \sqrt{(a/r)^2 - z}} dz \quad . \tag{115}$$

From a comparison of the right integrals in (115) and (112) it can be seen that, apart from the prefactor, both integrands differ only in the power of the term $(1-z)$ in the denominator. Hence, by means of partial integration it can be proved in a simple way that holds

$$I_3 = I_1 + \frac{\sqrt{\pi}\,\Gamma\left(1+\frac{k}{2}\right)}{2\Gamma\left(\frac{3+k}{2}\right)}\left(\frac{a}{r}\right)^{k+1}\left(1 - \frac{a^2}{r^2}\right)^{\frac{1-k}{2}} \;\Rightarrow\; I_2 = \frac{\sqrt{\pi}\,\Gamma\left(1+\frac{k}{2}\right)}{2\Gamma\left(\frac{3+k}{2}\right)}\left(\frac{a}{r}\right)^{k+1}\left(1 - \frac{a^2}{r^2}\right)^{\frac{1-k}{2}} \quad . \tag{116}$$

Substituting the results from Eqs. (113) and (116) into Eqs. (110) and (111) yields the displacement components sought outside the loading area:

$$\overline{u}_x(r,\theta) = \frac{\pi q_0 a^{1-k} c_0^k}{(k+1) E_0}\left(\frac{a}{r}\right)^{1+k}\left[(H+K)\,{}_2F_1\left(\frac{1+k}{2}, \frac{1+k}{2}; \frac{3+k}{2}; \frac{a^2}{r^2}\right) + (H-K)\left(1 - \frac{a^2}{r^2}\right)^{\frac{1-k}{2}} \cos 2\theta\right] , \tag{117}$$

$$\overline{u}_y(r,\theta) = \frac{\pi q_0 a^{1-k} c_0^k}{(k+1) E_0}(H-K)\left(\frac{a}{r}\right)^{1+k}\left(1 - \frac{a^2}{r^2}\right)^{\frac{1-k}{2}} \sin 2\theta \quad . \tag{118}$$